\documentclass[11pt,titlepage]{article}
\usepackage{graphicx}
\usepackage{rotating}
\usepackage{bm}
\usepackage{amsmath}
\usepackage{amssymb}
\usepackage{color}
\usepackage{setspace}
\usepackage[round,numbers,sort&compress]{natbib} 
\usepackage[margin=1in]{geometry}

\def\micron{$\mu$m}
\def\degree{$^{\rm o}$C}
\def\pers{ s$^{-1}$}
\def\permin{ min$^{-1}$}

\def\pombe{\textit{S.~pombe}}
\def\ttc{\langle \tau_c \rangle}
\def\meanl{\langle L \rangle}

\title{Contributions of microtubule dynamic instability and rotational
  diffusion to kinetochore capture}

\author{R. Blackwell, O. Sweezy-Schindler, C. Edelmaier,
  Z. R. Gergely, P. J. Flynn, S. Montes,\\ A. Crapo, A. Doostan,
  J. R. McIntosh, M. A. Glaser, and M. D.  Betterton}
\date{}

\begin{document} 
\maketitle

\abstract{ Microtubule dynamic instability allows search and capture
  of kinetochores during spindle formation, an important process for
  accurate chromosome segregation during cell division.  Recent work
  has found that microtubule rotational diffusion about minus-end
  attachment points contributes to kinetochore capture in fission
  yeast, but the relative contributions of dynamic instability and
  rotational diffusion are not well understood.  We have developed a
  biophysical model of kinetochore capture in small fission-yeast
  nuclei using hybrid Brownian dynamics/kinetic Monte Carlo simulation
  techniques. With this model, we have studied the importance of
  dynamic instability and microtubule rotational diffusion for
  kinetochore capture, both to the lateral surface of a microtubule
  and at or near its end.  Over a range of biologically relevant
  parameters, microtubule rotational diffusion decreased capture time,
  but made a relatively small contribution compared to dynamic
  instability. At most, rotational diffusion reduced capture time by
  25\%.  Our results suggest that while microtubule rotational
  diffusion can speed up kinetochore capture, it is unlikely to be the
  dominant physical mechanism for typical conditions
    in fission yeast. In addition, we found that when microtubules
  undergo dynamic instability, lateral captures predominate even in
  the absence of rotational diffusion. Counterintuitively, adding
  rotational diffusion to a dynamic microtubule increases the
  probability of end-on capture.

  }

\clearpage

\section*{Introduction}

Cell division is essential to the propagation of life. For a cell to
divide successfully, each daughter cell must inherit the correct
genetic material.  In eukaryotes, segregation of duplicated
chromosomes is performed by the mitotic spindle, a cellular machine
composed of microtubules (MTs) and their associated proteins
\cite{mcintosh12}.  Specialized sites on the chromosomes called
kinetochores (KCs) attach to spindle MTs, and these KC-MT attachments
are necessary for proper chromosome segregation.  Understanding KC
capture by spindle MTs and the subsequent chromosome movements is
challenging because the process depends on multiple overlapping
mechanisms \cite{cottingham97,goshima03,grishchuk06}, including the
action of multiple KC-associated motors and highly dynamic MTs that
maintain KC attachment during significant MT turnover.  Numerous
proteins localize to MT plus ends and KCs, but the roles of these
different proteins are not yet clear
\cite{schroer01,garcia02a,akhmanova08}.  Problems in kinetochore-MT
attachment and chromosome segregation can lead to aneuploidy, which is
associated with birth defects and cancer progression
\cite{duesberg06}.

The discovery of MT dynamic instability 30 years ago
\cite{mitchison84} led to the proposal that MT search and capture is
the primary mechanism of initial KC-MT attachment in mitosis. In this
picture, dynamic MTs grow in different directions from centrosomes and
make end-on attachments with KCs
\cite{mitchison85,hill85,holy94,heald15}. Perturbations to MT dynamics
are predicted to have significant effects on KC capture \cite{holy94},
suggesting that dynamic instability is a key component of any KC
capture model. Search and capture has been directly observed in large
cells \cite{rieder90}. However, the simplest search-and-capture
mechanism does not appear rapid enough to capture multiple chromosomes
quickly enough to match measured time in mitosis.  Extensions to the
search-and-capture mechanism that can make KC capture more rapid in
large cells or cell extracts include KC diffusion
\cite{holy94,paul09}, MT growth that is spatially biased toward
chromosomes \cite{carazo-salas99,wollman05,oconnell09}, chromosome
spatial arrangements and rotation \cite{paul09,magidson11,magidson15}
and KC-initiated MTs that can interact with searching MTs
\cite{witt80,kitamura10,paul09}. KCs in human cells change size and
shape during mitosis, which can both speed up capture and minimize
errors \cite{magidson15}.

\begin{figure*}[t]
  \centering
  \includegraphics[width=1.0 \textwidth]{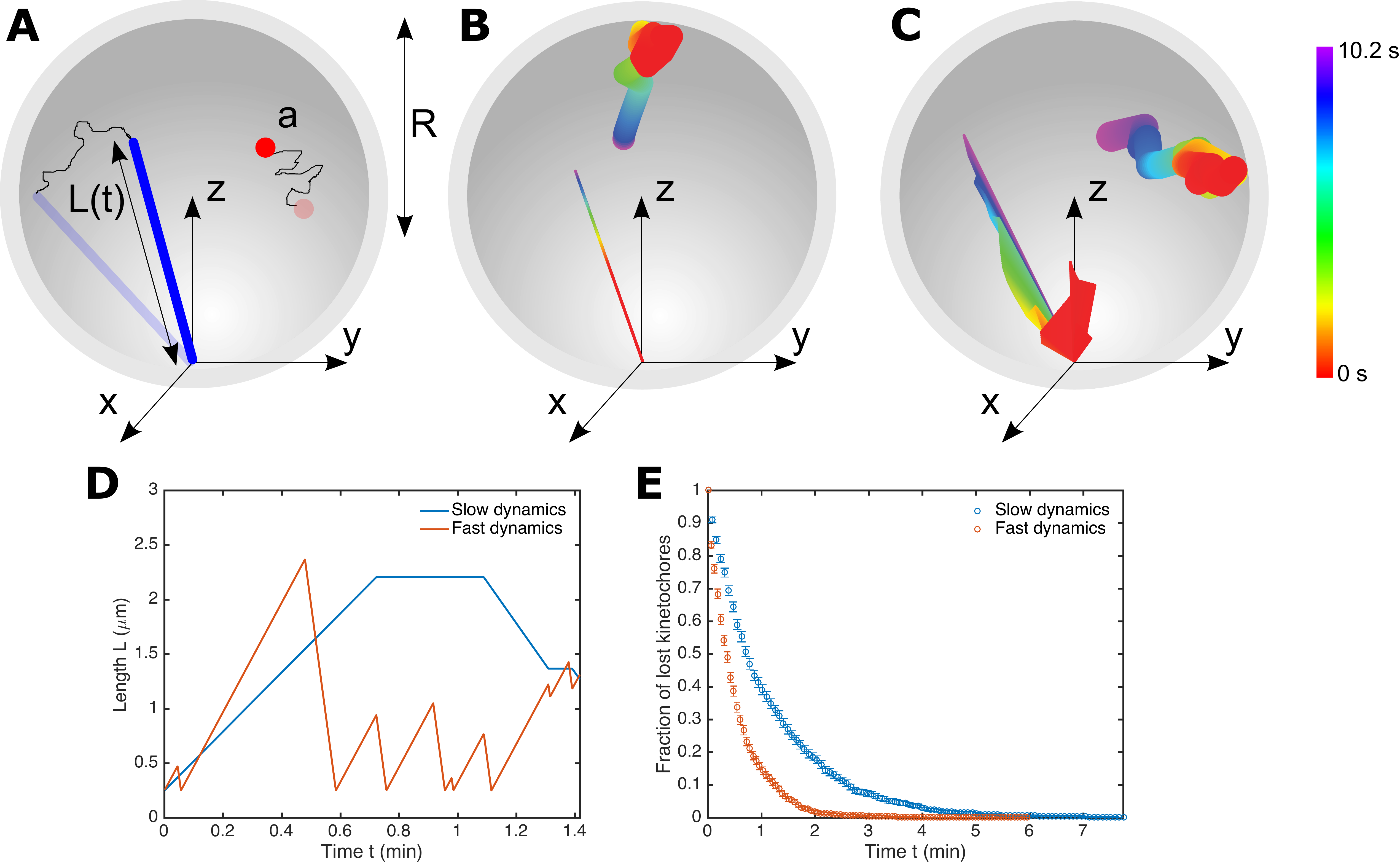}
  \caption {\footnotesize Kinetochore capture model overview. (A)
    Model schematic.  A dynamic microtubule (blue) undergoes dynamic
    instability and rotational diffusion about its minus-end
    attachment point; a kinetochore (red) undergoes translational
    diffusion within the spherical fission-yeast nucleus (gray).
    (B,C) Images created from simulation data showing microtubule
    (line) and kinetochore (circle) dynamics. Color indicates time,
    where red reflects an early position, purple a late one. (B) Model
    with microtubule dynamic instability but no rotational
    diffusion. (C) Model with dynamic instability and microtubule
    rotational diffusion.  (D) Microtubule length and (E) fraction of
    lost kinetochores versus time for slow (blue) and fast (orange)
    models with the reference parameters (table
    \ref{table:kc_cap_params}). }
  \label{fig:kc_capture}
\end{figure*}

KC capture may occur differently in smaller cells. Lateral KC contacts
have been directly observed both in budding yeast \cite{tanaka05} and
fission yeast \cite{kalinina12}.  Recent work on the small cells of
fission yeast found that lateral KC attachment to MTs that
rotationally diffuse about their attachment points at the spindle-pole
bodies (SPBs, the yeast centrosomes) enabled rapid KC capture, even
for relatively less dynamic MTs
\cite{kalinina12}. Here we use the term rotational
  diffusion for diffusive movements about MT minus-end attachment
  points, termed pivoting in previous work.  Based on experiments and
a biophysical model, this work concluded that MT rotational diffusion
was the primary determinant of the time to capture lost KCs in these
cells.  This important finding suggests that MT rotational diffusion
may significantly contribute to efficient KC capture, an effect that
has been neglected previously. Since initial KC captures are typically
lateral rather than end-on
\cite{rieder90,tanaka05,kitajima11,magidson11,kalinina12,magidson15},
MT rotational diffusion about minus-end attachment points could be an
important determinant of the capture time.

We have sought to evaluate the relative importance of MT dynamic
instability versus MT rotational diffusion to KC capture.  Previous
theoretical work has focused either on MT dynamic instability
\cite{hill85,holy94,paul09,wollman05,gopalakrishnan11,magidson15} or
rotational diffusion \cite{kalinina12}; as a result, this work has
been unable to compare the two mechanisms and determine their relative
importance. One model examined the relative contributions of dynamic
instability and rotational diffusion for a single parameter set and
found that turning off rotational diffusion caused a modest increase
in the mean capture time \cite{cojoc16}.  Because MT
rotational diffusion can lead to significant increases in the
effective volume searched by a single MT, search-and-capture models
that neglect MT rotational diffusion could lead to incorrect
conclusions.

We extended the Kalinina \textit{et al}.~model of fission-yeast KC
capture \cite{kalinina12} to include both MT rotational diffusion and
dynamic instability in order to gauge their relative importance to KC
capture. We modeled both relatively slow dynamic instability (using
measured parameters \cite{kalinina12}) and faster dynamic instability
based on our measurements of single mitotic MTs in fission yeast.
Rotational diffusion typically gave only a modest speed-up of KC
capture in our model. For slow dynamics, rotational diffusion
decreased the KC capture time by up to 25\%, while for faster MT
dynamics, rotational diffusion caused at most a 16\% decrease in the
capture time.  Our results suggest that the capture time in fission
yeast is primarily determined by MT search and capture.  We also found
that while lateral captures are typical even in the model with no
rotational diffusion, including MT diffusion made end-on attachments
more likely. This occurred because rotational diffusion sweeps the tip
of the MT through space, increasing the volume searched by the
tip. Our findings suggest that associating search and capture with
end-on attachment and rotational diffusion with lateral attachment is
oversimplified.

\section*{Materials and methods}

\subsection*{Kinetochore capture model}

We developed a computational model of KC capture in fission yeast that
includes the key physical effects of MT dynamic instability and
rotational diffusion.  Capture occurs within the spherical nucleus of
radius $R = 1.5\ \mu$m. A dynamic MT has its minus end attached to the
SPB (which is fixed for this study), rotationally diffuses, and has a
length $L(t)$ that changes with time due to dynamic instability
(fig.~\ref{fig:kc_capture}, \ref{suppfig:diff_verification}).
Simultaneously, a spherical KC of radius $a=100$ nm diffuses in the
nucleus \cite{kalinina12}. We assumed the fission yeast KC size
remained constant during mitosis, in contrast to recent work on human
KCs \cite{magidson15}.  KC capture occurs when the KC contacts the MT,
either at its end or along its lateral wall.

The simulations used a hybrid Brownian dynamics-kinetic Monte Carlo
scheme approach based on our previous work
\cite{gao15a,gao15,kuan15,blackwell16}.  Brownian dynamics model the
diffusive random motion of MTs and KCs; kinetic Monte Carlo models the
stochastic MT dynamic instability (Supporting Material).

\subsubsection*{Slow and fast microtubule dynamics}

We studied two MT dynamic instability models that represent relatively
slow and fast dynamics (Supporting Material,
fig.~\ref{fig:kc_capture}, table \ref{table:kc_cap_params}).  Kalinina
\textit{et al}.~found that MT dynamics were relatively slow and MTs
spent most of their time paused \cite{kalinina12}. Therefore, we
modeled MTs with growing, shrinking, or paused states; the
fixed-length paused state is an intermediate between the growing and
shrinking states (Supporting Material).  In other work on single
fission yeast mitotic \cite{sagolla03} and meiotic \cite{cojoc16} MTs
and our measurements, MT dynamics were faster and pausing was rarely
seen. We modeled this with growing and shrinking states only, where
catastrophe is the transition from growing to shrinking, and rescue
the transition from shrinking to growing.

In our model, MTs in the growing or shrinking state increase or
decrease in length at the constant speed $v_g$ or $v_s$.  For MTs that
are not interacting with the nuclear envelope boundaries, the state
switching frequencies are constant in time.  Any shrinking MTs that
reach the minimum length of 4$\sigma_{MT}$ = 100 nm switch to the
growing state (Supporting Material).

\subsubsection*{Microtubule interactions with the nuclear envelope}

MTs that touch the nuclear envelope experience steric forces and
torques from the interaction of the MT tip with the envelope
(Supporting Material). The torque can cause the MT tip to slip along
the edge of the envelope, reorienting the MT, as has been measured and
modeled previously for MTs interacting with microchamber boundaries
\cite{laan12,pavin12,ma14}.  In addition, MTs that grow into a
boundary exhibit increased catastrophe frequency \cite{tischer09}. The
force component along the MT long axis increases the catastrophe
frequency, as measured previously \cite{dogterom97,janson03}. By
combining these previous measurements of the force dependence of MT
growth speed with the growth speed dependence of the catastrophe time,
we wrote the catastrophe frequency
$f_c(F_{||}) = f_c \exp(\alpha F_{||})$, where $F_{||}$ is the
component of the steric force along the MT long axis,
$\alpha$ is the force sensitivity of catastrophe
  (table \ref{table:kc_cap_params}), and $f_c$ is the zero-force
catastrophe frequency (or the analogous grow-to-pause frequency in the
model with pausing).

\subsubsection*{Initial conditions and measurements}

We began simulations with a single MT of length 4$\sigma_{MT}$ (100
nm) placed at a random angle subject to the requirement that the MT
was not initially interacting with the nuclear envelope. We inserted a
KC at a random position uniformly sampled within the simulation
volume, with the requirement that the KC was not initially interacting
with either the MT or the nuclear envelope. We ran the simulation
until the KC collided with the MT (either laterally or end-on), which
defined a capture. For each parameter set, we repeated simulations
2000-5000 times to determine the distribution of capture times, shown
as the fraction of lost KCs as a function of time in
figs.~\ref{fig:kc_capture}D, \ref{suppfig:kalinina_verification}. From
these data, we computed the mean capture time $\ttc$.

We defined reference parameter sets and wide parameter ranges around
the reference for both the slow and fast dynamic instability models
(table \ref{table:kc_cap_params}). For both slow and fast models, we
performed simulations of dynamic instability with no MT rotational
diffusion, and dynamic instability plus MT rotational diffusion.  To
connect to previous search-and-capture models, MTs that shrink to the
minimum length re-enter the growing state with a new random
orientation.

\subsection*{Experimental methods}

To understand the difference between previous measurements of
fission-yeast mitotic MT dynamics that were relatively slow
\cite{kalinina12} or fast \cite{sagolla03}, we measured MT dynamic
instability in \pombe.  To facilitate these measurements, we used a
strain with temperature-sensitive inactivation of kinesin-5 motors
(\textit{cut7-24} in fission yeast) and low-level fluorescent tagging
with \textit{mCherry-atb2} \cite{yamagishi12} (Supporting Material,
table \ref{table:strain_table}).  Cells carrying the \textit{cut7-ts}
allele are unable to form bipolar mitotic spindles at restrictive
temperature (36-37$^o$C) \cite{hagan90}. The cells instead form
monopolar mitotic spindles, in which individual fluorescently labeled
mitotic MTs can be imaged (fig.~\ref{fig:experiment_data_kc})
\cite{costa13}.

\subsubsection*{Measurement of labeled tubulin fraction}

\begin{figure*}[t]
  \centering
  \includegraphics[width=1.0 \textwidth]{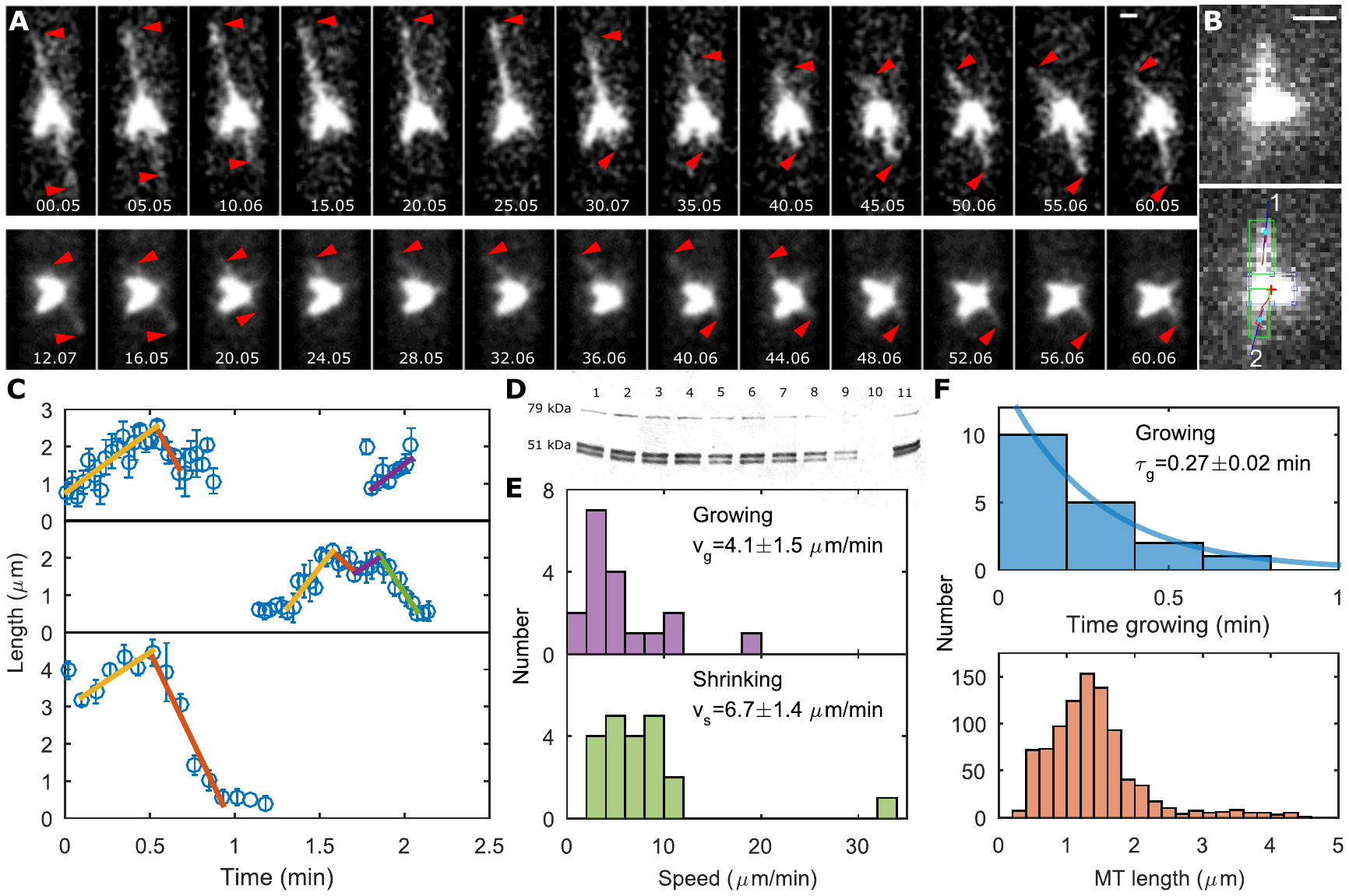}
  \caption {\footnotesize Measurements of single MT dynamics and
    length in monopolar mitotic spindles. (A) Image sequences. Images
    are pixel-interpolated maximum intensity projections of monopolar
    spindles with mCherry-labeled MTs. Red arrowheads indicate single
    MTs, time is shown in sec, and the scale bar is 1 \micron. (B)
    Example raw maximum-intensity-projected image (top) and analyzed
    image (lower). The red cross indicates the center of brightness of
    the SPB, green boxes the regions analyzed near MT tips, blue lines
    the fits to the MT backbone positions, and blue dots the fits to
    the MT tip positions.  The scale bar is 1 \micron. (C) Example
    traces of MT length versus time (points) and fits to growing and
    shrinking events (lines). (D) Western blot used to quantify
    fraction of fluorescently labeled tubulin. The gel
      was loaded with serial dilutions from 100--10\% with 100\%
      repeated in lane 11: lane 1=100\%, lane 2=90\%, lane 3=80\%,
      lane 4=70\%, lane 5=60\%, lane 6=50\%, lane 7=40\%, lane
      8=30\%, lane 9=20\%, lane 10=10\%, lane 11=100\%. (E)
    Histograms of growth speed and shrinking speed. (F) Histograms of
    growth time and MT length. }
  \label{fig:experiment_data_kc}
\end{figure*}

We performed immunoblots using the TAT-1 tubulin antibody on \pombe\
cell lysate with serial dilutions ranging from 100\%-10\% of the
original cell suspension concentration (Supporting Material) and
scanned the bands for analysis. The two lower, darker bands
corresponded to $\alpha$-tubulin-1 (nda2) at 51 kDa and
$\alpha$-tubulin-2 (atb2) at 50 kDa, while the fainter third band
corresponded to mCherry-atb2 at 79 kDa
(fig.~\ref{fig:experiment_data_kc}). To analyze the scanned images, we
inverted the images so that labeled regions corresponded to high
intensity, drew equally-sized regions of interest around each band, and determined
the average pixel intensity in each region. From this,
we determined the fraction of intensity in the mCherry-atb2 band relative to the
total. Each lane had similar ratios (data not shown), and we averaged
the results for each lane.

\subsubsection*{Cell preparation and confocal imaging}
We grew cells using standard techniques (Supporting Material) and
cultured them on glass-bottomed dishes at 36$^o$C for 2--4 hrs to
allow monopolar spindles to form. The dishes were transferred to the
microscope in less than 60 sec to prevent the monopolar spindles from
becoming bipolar. Images were taken on an spinning disk (Yokogawa,
Musashino, Japan) Nikon Eclipse Ti inverted confocal microscope
(Nikon, Tokyo, Japan) with a 100X, 1.4 NA Plan Apo oil-immersion
objective, an iXon Ultra 897 EM-CCD camera (Andor, Belfast, United
Kingdom) and a TIZSH Stage Top incubator (Tokai Hit, Fujinomiya,
Japan) warmed to 36$^o$C.  Three-dimensional images were obtained with
an EM Gain of 300, and an exposure time per plane of 40 to 150 msec
with 595 nm laser illumination for each of 5 focal planes separated by
500 nm in z, and subsequent stacks are separated by 4--6 sec.  Images of
fig.~\ref{fig:experiment_data_kc}A are displayed as pixel-interpolated
maximum-intensity projections.

\subsubsection*{Image and data analysis}

In our images of monopolar spindles, we were able to observe dynamic
MTs (fig.~\ref{fig:experiment_data_kc}).  We quantified the dynamic
instability of mitotic MTs by determining the location of the SPBs and
the tip of each emanating MT with image analysis software adapted from
TipTracker \cite{demchouk11,prahl14}.  To identify the center of the
monopolar spindle, we noted that near the two spindle-pole bodies
(SPBs), many short and overlapping MTs produce a bright fluorescent
region. When using this software, one selects a rectangle around this
region, and our program then computes the center of intensity within
that region to estimate the SPB location and determines the length of
each MT in a given frame by assuming the MT formed a line between the
estimated SPB location and the MT tip position. Visual inspection of
each frame confirmed that MTs in the monopolar spindles formed in
these cells were typically straight lines emanating from the bright
spindle center region (fig.~\ref{fig:experiment_data_kc}B).  Each
frame is analyzed separately, and the MT length data stored for
analysis.  The software allows us to make MT length and angle
measurements with subpixel resolution and quantify the lengths and
dynamics of mitotic MTs in monopolar spindles.

We identified growth and shrinking events by comparing the movies and
plots of MT length versus time to identify starting and ending times
of events. We then performed weighted least-squares linear fits to the
MT length versus time during each event
(fig.~\ref{fig:experiment_data_kc}). Growth and shrinking speeds were
the slopes determined from the fits, and the catastrophe and rescue
times were the duration of the events before a switch.

\begin{table*}[t]
\footnotesize
  \centering
  \setlength{\tabcolsep}{6pt}
  \begin{tabular}{llllp{3cm}}
    \textbf{Parameter} & \textbf{Symbol} & \textbf{Reference value} & \textbf{Range} & \textbf{Notes} \\
    Nuclear envelope radius  & R & 1.5 $\mu$m&-- &
                                                   \citet{kalinina12} \\
    KC diameter  & $\sigma_{KC}$ & 200 nm &-- & \citet{ding93} \\
    MT diameter  & $\sigma_{MT}$ & 25 nm &-- & \citet{alberts08} \\
    MT angular diffusion coefficient  & $D_{\theta}$ & --
                                         & Varies with MT length &
                                                                   \citet{kalinina12},
    Supporting Material\\
    Force-induced catastrophe constant  & $\alpha$ & 0.5 pN$^{-1}$ &--
                                                 & \citet{dogterom97}
                                                   and \citet{janson03} \\
  
    \hline    
    \textbf{Slow dynamic instability model} & & & & \\
    Growth speed & $v_{g}$ & 2.7  $\mu$m\permin & 0.7--11  $\mu$m\permin & \citet{kalinina12} \\
    Shrinking speed  & $v_{s}$ & 3.8  $\mu$m\permin & 1--16
                                                      $\mu$m\permin
                                                 &
                                                   \citet{kalinina12} \\ 
    Grow-to-pause frequency  & $f_{+0}$ & 1.8 \permin & 0.4--7.1 \permin &  \citet{kalinina12} \\
    Shrink-to-pause frequency  & $f_{-0}$ & 2.53 \permin & 1.2--5 \permin & \citet{kalinina12}\\
    Pause-to-shrink frequency  & $ f_{0-}$ & 0.49 \permin & 0.2--1 \permin & \citet{kalinina12} \\
    Pause-to-grow frequency  & $ f_{0+}$ & 0 \permin & --& Transitions
                                                           from pausing
                                                           to growing
                                                           appeared infrequent
                                                           in \citet{kalinina12}\\
    \hline    
    \textbf{Fast dynamic instability model} & & & & \\
    Growth speed  & $v_{g}$ &  4.1 $\mu$m\permin & 1--10 $\mu$m\permin & This work \\
    Shrinking speed  & $v_{s}$ &   6.7 $\mu$m\permin & 5--25 $\mu$m\permin &
                                                                             This
                                                                             work and \citet{sagolla03}\\
    Catastrophe frequency  & $f_{\rm cat}$ &  3.7 \permin & 1--8 \permin &
                                                                           This
                                                                           work \\
    Rescue frequency  & $f_{\rm res}$ &  0.175 \permin & 0--8 \permin & This work\\
  \end{tabular}
  \caption{\footnotesize Model parameter values. }
  \label{table:kc_cap_params}
\end{table*}

\section*{Results and discussion}

\subsection*{Experimental results}

Kalinina \textit{et al}. \cite{kalinina12} found that fission yeast
mitotic MTs on average spent 75\% of their time in a paused state and
had lifetimes of 3 min, growth speed $v_g = 2.7$ \micron \permin, and
shrinking speed $v_s = 3.8$ \micron \permin. These results differed
from the results of Sagolla et al. \cite{sagolla03}, who observed
highly dynamic polar MTs in early mitosis (before spindle formation),
with lifetimes of seconds and a shrinking speed $v_s = 20$ \micron
\permin, and more recent work on meiotic MTs \cite{cojoc16}. These
differences in lifetime and dynamics could be related to the stage of
mitosis (before and after spindle formation), the number of MTs per
bundle, and/or to the fraction of fluorescent tubulin in the cells
(which affects MT dynamics \cite{snaith10}). The differences are most
likely due to MT bundling, which alters MT dynamics
\cite{bratman07,bratman08}. Since KC capture could occur either by
single MTs that are more dynamic or bundled MTs that are more stable,
we undertook additional measurements of mitotic MT dynamic instability
in \pombe.

We adapted the strategy of Costa \textit{et al}. \cite{costa13}, who
used temperature-sensitive inactivation of the kinesin-5 motor
(\textit{cut7-24}) to obtain cells stably arrested in a monopolar
state. Fission yeast carrying the \textit{cut7-ts} allele arrest in
early mitosis at restrictive temperature (36-37$^o$C), because bipolar
spindles cannot form when cut7p is inactive \cite{hagan90}. These
cells instead form monopolar mitotic spindles, in which individual
fluorescently labeled mitotic MTs can be imaged and tracked
(fig.~\ref{fig:experiment_data_kc}A, B) \cite{costa13}.

To measure MT length, we adapted the TipTracker algorithm
\cite{demchouk11,prahl14} to measure MT lengths in monopolar spindles
(Methods, fig.~\ref{fig:experiment_data_kc}B).  Brighter MTs showed
lengths that were more stable in time, and dimmer MTs showed more
rapid dynamics (fig.~\ref{fig:experiment_data_kc}A). We identified the
dimmer, more dynamic MTs as single MTs, as in previous work
\cite{costa13}. We analyzed 20 MTs from 15 cells that showed low
intensity compared to other MTs in the same cell and relatively fast
dynamics.  We determined growth and shrinking events and their
associated speeds and times (fig.~\ref{fig:experiment_data_kc}C,
Methods).

Fluorescent-protein fusions to tubulin in fission yeast can alter MT
dynamics \cite{snaith10}. We studied fission yeast carrying
\textit{mCherry-atb2} that is an additional copy of this
$\alpha$-tubulin gene under a weak promoter \cite{yamagishi12}. We
used Western blotting to determine the fraction of tubulin our cells
that was fluorescently tagged (fig.~\ref{fig:experiment_data_kc}D,
Methods) and found a low fraction of 8.8\% $\pm$ 0.5\% labeled
$\alpha$-tubulin.

These MTs had a median growth speed $v_g= 4.1 \pm 1.5$ \micron
\permin\ and shrinking speed $v_s = 6.7\pm 1.4$ \micron \permin\
(fig.~\ref{fig:experiment_data_kc}E).  The distribution of times in
the growing state appeared exponential with a characteristic time
$\tau_g = 0.27\pm 0.02$ min, implying a catastrophe frequency of 3.7
\permin\ (fig.~\ref{fig:experiment_data_kc}F).  Measuring the rescue
frequency was challenging, because it was difficult to distinguish
rescue from complete shrinkage followed by regrowth of a different MT
in the same area.  We saw one possible rescue event, which gave a
bound $f_r \le 0.175$ \permin.

Our data were collected at 37\degree, and MT dynamic instability is
sensitive to temperature \cite{fygenson94}. However, Kalinina
\textit{et al}.~found relatively little change in fission-yeast MT
dynamic instability parameters between 24\degree\ and 32\degree\
\cite{kalinina12}, suggesting that \pombe\ mitotic MT dynamic
instability may not vary markedly with temperature in this
regime. Another possible complication is that MT
  dynamics might be altered in the monopolar spindle system where
  kinesin-5 is inactive. Previous work in budding yeast found that
  kinesin-5s promote disassembly of longer MTs
  \cite{gardner08}. However, since we found similar MT mean lengths to
  those of Kalinina et al.~\cite{kalinina12}, it appears that our
  measurements do not show a strong bias toward longer MTs due to loss
  of kinesin-5-associated depolymerization activity.

Pooled length measurements of MTs we identified as single MTs
(including all measurements, not just points identified as
growing/shrinking events) had a median $\meanl = 1.31 \pm 0.02$
\micron\ (fig.~\ref{fig:experiment_data_kc}F).  We compared this to
the predicted mean length for dynamic instability in an infinite
volume (neglecting boundary effects): in the bounded growth regime
$\meanl= v_g v_s/(v_s f_c - v_g f_r)$ \cite{dogterom93}. If we
estimate the mean length using our median dynamic instability
parameters with the upper bound on the rescue frequency of
$f_r \le 0.175$ \permin, we predict $\langle L \rangle = 1.14$
\micron, while using $f_r = 0$ gives $\langle L \rangle = 1.11$
\micron. There is little difference between these values, showing that
the mean MT length is primarily determined by the growth speed and
catastrophe frequency in this low-rescue-frequency regime. This
typical MT length estimated from the dynamic instability parameters is
consistent with our direct measurements of MT length.

\subsection*{Model results}

\subsubsection*{Kinetochore capture by single microtubules with fast or
  slow dynamic instability}

Since KC capture could occur either by single MTs that are more
dynamic or bundled MTs that are more stable, we sought to understand
how these different MT arrangements affect KC capture in fission
yeast.  We studied two reference parameter sets that we denote
\textit{slow} (based on the Kalinina \textit{et al}.~measurements) and
\textit{fast} (based on our measurements).  In the slow model, MTs can
be growing, shrinking, or paused.  The reference parameter values
(table \ref{table:kc_cap_params}) were taken from Kalinina et al.,
\cite{kalinina12} assuming that transitions from pausing to growing
did not occur. In the fast model, we considered only growing and
shrinking states with reference parameter values determined from our
measurements.  Examples of the resulting MT length as a function of
time and the dynamics of KC capture are shown in figure
\ref{fig:kc_capture}. For each model, we varied parameters around the
reference values by factors of 4--20 (table \ref{table:kc_cap_params})
so that we could study the dependence of the capture time on parameter
values. We then studied both slow and fast models with and without MT
rotational diffusion, to determine the contributions of dynamic
instability alone and dynamic instability with rotational diffusion to
the capture time.

Kalinina \textit{et al}.~found that 3 polar MTs were typically visible
during their KC capture experiments, and that the time of KC capture
was sensitive to MT number \cite{kalinina12}. Here we determined
capture times for a single MT. Our model results for capture by a
single MT and slow dynamic instability agree well with those of
Kalinina \textit{et
  al}.~(fig.~\ref{suppfig:kalinina_verification}). Studying capture by
single MTs allowed us to focus on the effect of dynamic instability
parameters and rotational diffusion.


\begin{figure*}[t]
  \centering
  \includegraphics[width=1.0 \textwidth]{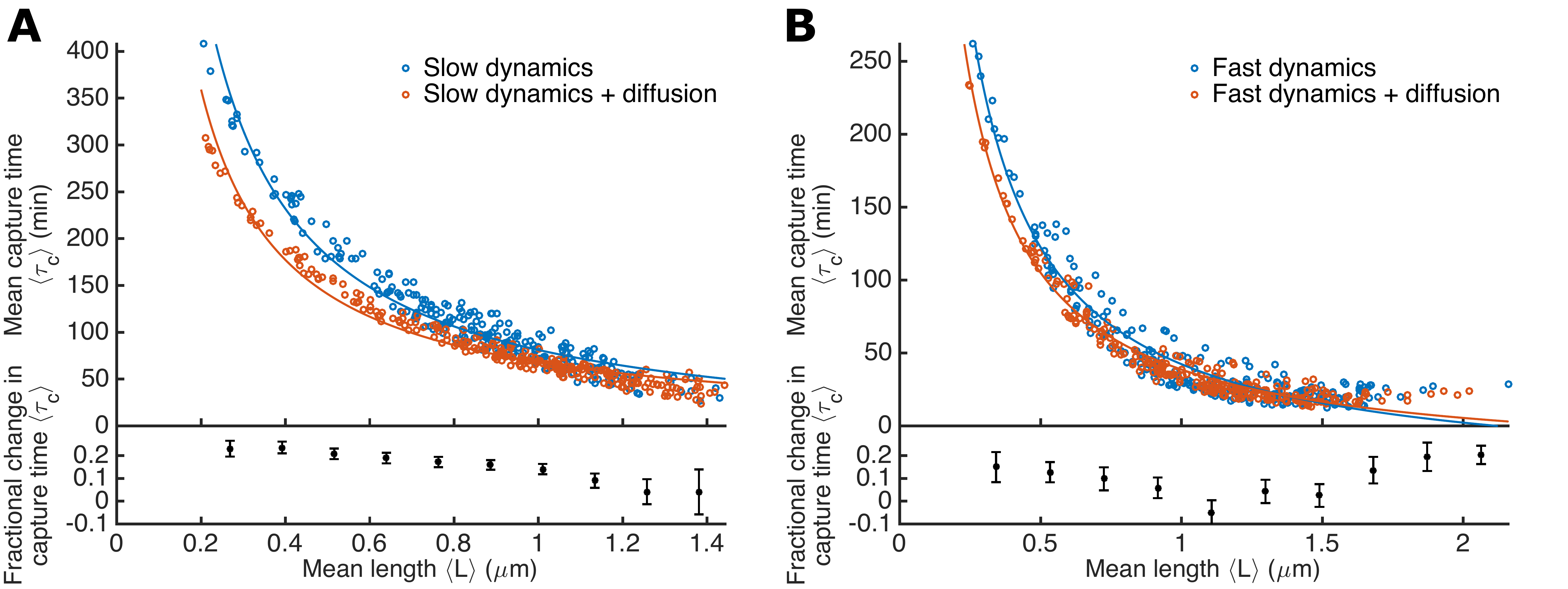}
  \caption {\footnotesize Dependence of capture time on MT length and
    rotational diffusion.  (A) Slow model.  (B) Fast model.  Points
    are results simulations with all dynamics parameters varied for
    search and capture only (blue) and search and capture with
    microtubule rotational diffusion (orange).  Lines are fits to
    $\ttc = A \meanl^{-1}+B$.  Insets, fractional decrease in capture
    time when rotational diffusion is added to the
    model. Error bars are determined using standard
      error propagation techniques from the standard error of the mean
      capture time in each bin of simulations. }
  \label{fig:ttc_vs_l}
\end{figure*}

\subsubsection*{Parametric study
  using polynomial chaos expansion}

To understand how the mean capture time $\ttc$ and the mean MT length
$\meanl$ depend on model parameters (table \ref{table:kc_cap_params}),
we used polynomial chaos (PC) expansion \cite{Ghanem03,Xiu02}, a
widely used technique for uncertainty quantification. PC expansion is
a type of spectral method in which we represent $\ttc$ and $\meanl$ as
functions of the dynamic instability parameters in a high-order,
multivariate, orthogonal polynomial basis, here of Legendre type. This
allowed us to use a relatively small number of simulations,
corresponding to random samples of input parameters generated
uniformly over their allowed ranges (table \ref{table:kc_cap_params}),
to accurately approximate $\ttc$ and $\meanl$ over the full
multidimensional parameter space \cite{doostan11a,
  hampton15a,hadigol15}. We then used this expansion to perform global
and local sensitivity analysis to determine which dynamic instability
parameters are of most importance to $\ttc$ and $\meanl$. The global
sensitivity analysis provides information on the importance of each
parameter in terms of its contribution to the overall solution
variability, while the local analysis identifies the dependence of the
solution on each parameter in the small neighborhood of its nominal
value. To construct the PC expansions, here of total degree three, we
used a regression approach based on $\ell_1$-minimization
\cite{doostan11a, hampton15a,hadigol15} using 250 randomly sampled
parameter sets, and the corresponding realizations of $\ttc$ and
$\meanl$.  In figs.~\ref{fig:ttc_vs_l} and
\ref{fig:l_vs_params}--\ref{fig:lcap_vs_l} below, each point
corresponds to one of the 250 parameter sets.

\subsubsection*{Effects of varying mean MT length}

The rotational diffusion coefficient of a rod varies as $L^{-3}$,
making MT length important for KC capture by a fixed-length MT
\cite{kalinina12}. We found that the MT mean length was also important
for KC capture with dynamic instability. Indeed, the primary
determinant of the capture time was the mean MT length
(fig.~\ref{fig:ttc_vs_l}), for all models studied. The capture time
decreased by approximately a factor of 10 as $\meanl$ increased from
short ($\lesssim 0.3$ \micron) to long ($\gtrsim 1.5$ \micron, half
the nuclear diameter).  A longer MT increases the effective number of
binding sites for a KC, lowering the capture time. If the number of
binding sites were the sole factor determining the rate of KC capture,
we would expect $\ttc \sim \meanl^{-1}$. The dependence in our model
is more complex, because the MT dynamics also change with $\meanl$.
Nevertheless, fitting the capture time as a function of the mean
length to the form $\ttc = A \meanl^{-1}+B$ (solid lines in
fig.~\ref{fig:ttc_vs_l}) gave reasonable agreement with our simulation
results.

For all MT lengths and both slow and fast dynamic instability, adding
MT rotational diffusion to a model with only dynamic instability
reduced the capture time. The contribution of rotational diffusion
depended on the MT dynamics: $\ttc$ decreased by 23\% on average (25\%
at most) for the slow model, and by 9\% on average (16\% at most) for
the fast model. The speed up due to rotational diffusion was larger
for slower MT dynamics and shorter MTs (fig.~\ref{fig:ttc_vs_l}). For
the most relevant MT lengths in fission yeast of around 1 \micron,
rotational diffusion shortened the capture time by at most
14\% for the slow model and 6\% for the fast model.
This suggests that while rotational diffusion does speed up KC
capture, it makes a relatively small quantitative contribution,
consistent with recent work \cite{cojoc16}.

\subsubsection*{Sensitivity analysis}

We performed sensitivity analysis to check how the capture time and MT
mean length vary with model parameters. For dynamic instability with
no boundary effects, we would expect a mean length of
$\meanl = v_g/f_{+0}$ in the slow model and
$\meanl = v_g v_s/(v_s f_c - v_g f_r) \approx v_g/f_c$ in the fast
model (\cite{dogterom93} and Supporting Material).  While interactions
of MTs with the nuclear envelope alter this relationship, we expected
that the MT mean length and therefore the capture time depend
primarily on the MT growth speed and the catastrophe frequency (or its
analogue in the slow model, the grow-to-pause frequency).

To test these relationships, we performed a global
  sensitivity analysis of the mean capture time and MT length to the
  dynamic instability parameters using the analysis of the variance of
  $\ttc$ and $\meanl$ based on the so-called Sobol' decomposition
  \cite{sobol90}, which we computed directly using the PC expansion
  (\cite{sudret08} and Supporting Material). The PC expansion gave low
  errors of a few percent (table \ref{table:pc_error}), indicating
  that it accurately describes the full simulation model. As expected,
  $\ttc$ and $\meanl$ are most sensitive to the growth speed and
  effective catastrophe frequency, and this dependence is not altered
  significantly by the addition of rotational diffusion to the model
  (table \ref{table:sobol_indices}). We also determined the local
  sensitivity of the capture time to these parameters (Supporting
  Material, fig. \ref{fig:ttc_vs_vg_fcat_fixed}).

\
\begin{figure*}[t]
  \centering
  \includegraphics[width=1.0 \textwidth]{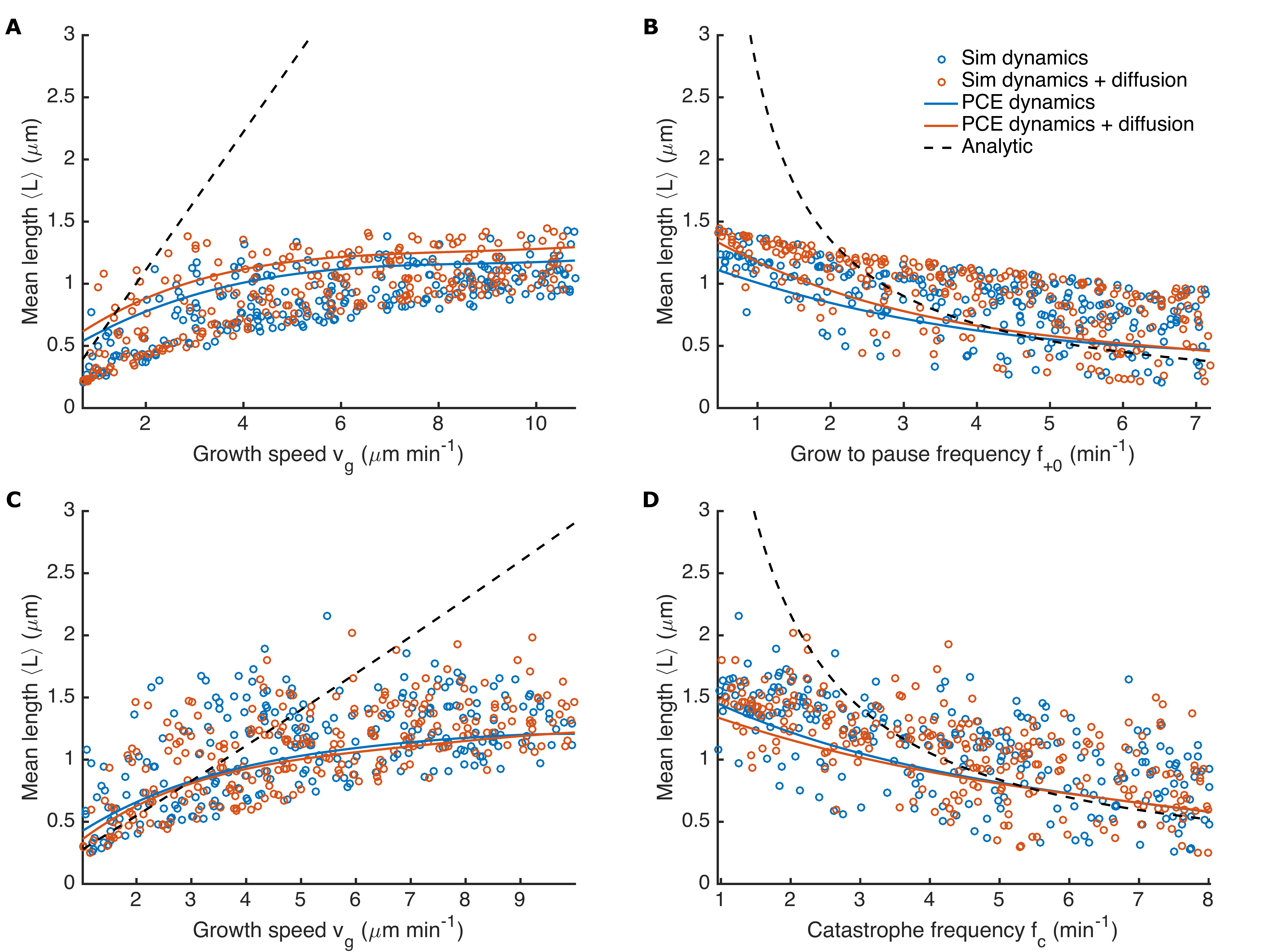}
  \caption {\footnotesize Variation of microtubule length with single
    parameters.  (A, B) Slow model. (C, D) Fast model.  Points
    represent individual simulations of model with dynamic instability
    only (blue) and model with dynamic instability and rotational
    diffusion (orange). Solid lines indicate a
      projection of the polynomial chaos expansion prediction
    in one dimension for variation of indicated
    parameter with other parameters at their reference values. Dashed
    line indicates predicted mean microtubule length for dynamic
    instability in the absence of boundary effects.}
  \label{fig:l_vs_params}
\end{figure*}

\subsubsection*{Variation with individual parameters}

To develop better intuition for how variation of
  individual parameters affects capture, we used both our individual
simulations and the PC expansion to study how the MT mean length and
capture time depend on variation of individual parameters
(figs.~\ref{fig:l_vs_params}, \ref{suppfig:l_ttc_vs_all_slow},
\ref{suppfig:l_ttc_vs_all_fast}). The solid lines are the predicted
dependence of $\meanl$ for the reference parameter set on the varied
single parameter from the PC expansion, while the dashed line is the
analytic prediction for $\meanl$ with unconstrained dynamic
instability. As expected, the analytic solution that neglects boundary
interactions matches the data and PC expansion well when $\meanl$ is
relatively small, but differs significantly where $\meanl \gtrsim 1$
\micron. The points show the behavior of individual
  simulations with widely varying parameters, illustrating the range
  of behavior of our model.

\begin{figure*}[t]
  \centering
  \includegraphics[width=1.0 \textwidth]{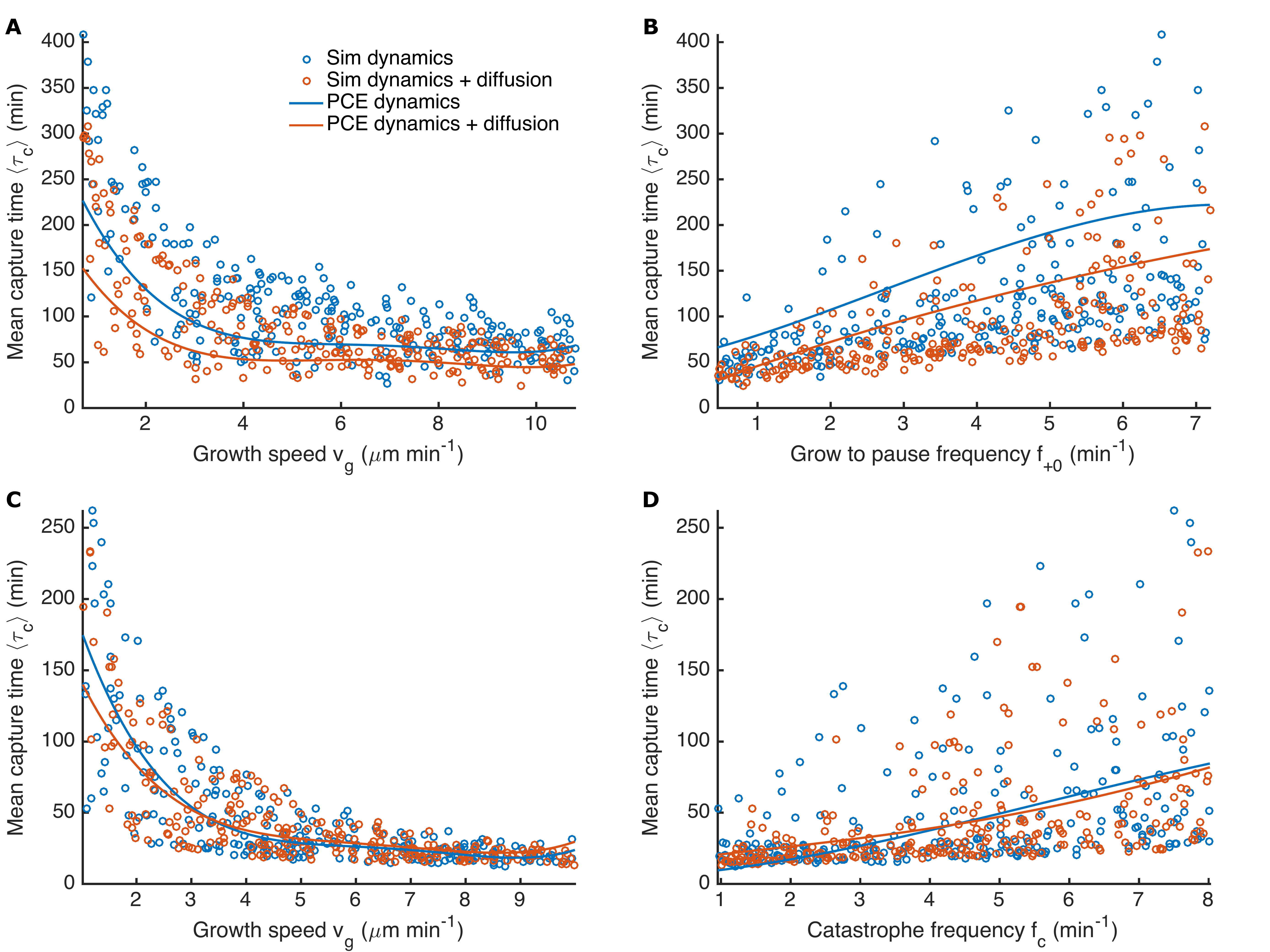}
  \caption {\footnotesize Variation of capture time with single
    parameters.  (A, B) Slow model. (C, D) Fast model.  Points
    represent individual simulations of model with dynamic instability
    only (blue) and model with dynamic instability and rotational
    diffusion (orange). Solid lines indicate a
      projection of the polynomial chaos expansion prediction
    in one dimension for variation of indicated
    parameter with other parameters at their reference values. }
  \label{fig:ttc_vs_params}
\end{figure*}

The variation of the capture time with individual parameters show
qualitatively similar dependence (figs.~\ref{fig:ttc_vs_params},
\ref{suppfig:l_ttc_vs_all_slow}, \ref{suppfig:l_ttc_vs_all_fast}).
Consistent with intuitive expectations, parameter
sets with higher growth speed and lower catastrophe frequency lead to
shorter capture times.  Growth speed is the main parameter affecting
capture time, causing it to vary by up to a factor of five, consistent
with its total Sobol' index (table
\ref{table:sobol_indices}). This makes sense because
MTs with higher growth speeds search a given direction more quickly
than slower MTs with the same orientation, resulting in lower capture
time; fast-growing MTs have also typically longer mean
lengths. Also as expected, catastrophe frequency
affects the capture time, because high catastrophe frequency tends to
reduce the mean length of MTs. However, MTs with low
catastrophe frequency typically reach the nuclear envelope and undergo
force-induced catastrophe, which limits the advantage gained by
lowering the catastrophe frequency.  The shrinking speed and rescue
frequency have little effect on the capture time, because they don't
significantly affect the mean MT length
(figs.~\ref{suppfig:l_ttc_vs_all_slow},
\ref{suppfig:l_ttc_vs_all_fast}). For simulations both with and
without MT rotational diffusion, varying these parameters alone varies
the capture time by $\lesssim 25$\%.

\subsubsection*{Lateral versus end-on capture}

\begin{figure*}[t]
  \centering
  \includegraphics[width=1.0 \textwidth]{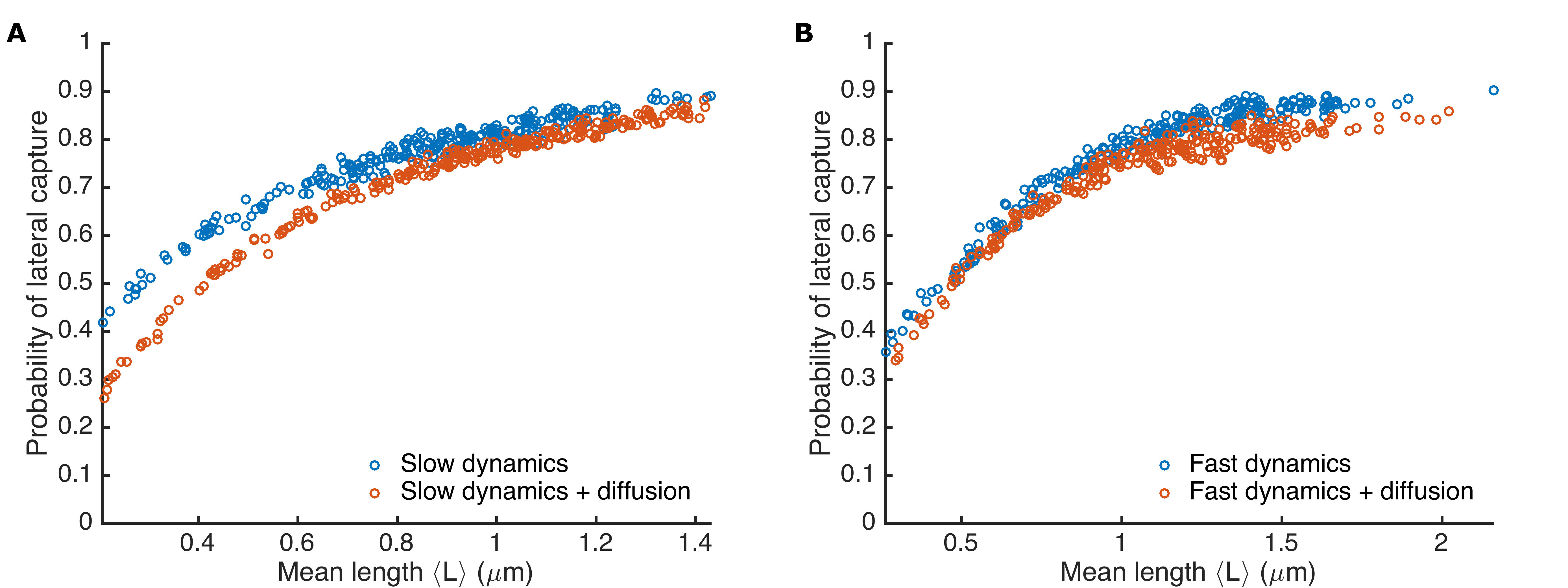}
  \caption {\footnotesize The fraction of captures
      that occur laterally as a function of mean microtubule
    length. (A) Slow model. (B) Fast model.  Points represent
    individual simulations of model with dynamic instability only
    (blue) and model with dynamic instability and rotational diffusion
    (orange).  }
  \label{fig:lcap_vs_l}
\end{figure*}

In fission yeast, KC capture occurs primarily via
end-on attachment, with about 75\% of captures
occurring $<$500 nm away from the MT tip
\cite{kalinina12}, while other other work both in
  large and small cells has found a high frequency of lateral
  attachments
  \cite{rieder90,tanaka05,kitajima11,magidson11,magidson15}.  In our
model, we classified attachments as end-on if the capture occurred
within 4.5$\sigma_{MT}$ ($\approx$ 113 nm) from the MT tip, and
lateral otherwise.  We found that lateral attachments are more likely
in our model, but the fraction of captures that occur laterally
depends on the mean MT length (fig.~\ref{fig:lcap_vs_l}).  Increasing
$\meanl$ increased the probability of lateral capture, because the
available MT surface area for binding laterally increases.

Remarkably, we found that lateral captures predominated even in the
model with dynamic instability only (no rotational diffusion). This
occurred for two reasons: first, interactions of MTs with the nuclear
envelope caused some MT reorientation in the absence of thermal
diffusion. However, we observed significant lateral attachment even
for relatively short MTs that didn't interact with the nuclear
envelope, because KC diffusion allowed lateral attachment even for
fixed-orientation MTs. 

Surprisingly, adding rotational diffusion to the model made end-on
attachments more likely: diffusion \textit{decreases} the probability
of lateral capture, an effect that was more noticeable for shorter
MTs. Rotational diffusion allows the MT to sweep through space,
increasing the volume searched by the MT. This increase in effective
volume searched is largest at the MT tip because the tip is furthest
from the pivot point at the SPB. This effect was more important for
shorter MTs where end-on attachments were more likely. 

It's not clear why our model found primarily lateral
  captures, while Kalinina et al. \cite{kalinina12} measured primarily
  end-on captures. Some difference would be expected due to the
  different definitions of end-on versus lateral attachment: Kalinina
  et al.~used a 500-nm distance cutoff that was limited by optical
  resolution. In addition, it is possible that because lost KCs were
  induced by a short cold treatment in these experiments, more of the
  KCs were close to the SPB and were captured by shorter MTs for which
  rotational diffusion is more important.

\section*{Conclusions}

The search-and-capture model has been considered the main mechanism by
which KCs are captured since the discovery of MT dynamic instability
\cite{mitchison84,hill85,holy94,heald15}. The original picture of
search and capture posited that MTs undergo dynamic instability
nucleated from centrosomes until they form end-on attachments with
KCs. Additional effects such as diffusion of KCs,
\cite{holy94,paul09}, a bias in MT growth toward chromosomes
\cite{carazo-salas99,wollman05,oconnell09}, MT nucleation from KCs
\cite{witt80,kitamura10,paul09}, spatial distribution and rotation of
chromosomes \cite{paul09,magidson11,magidson15}, and KC size decreases
after capture \cite{magidson15} can make search and capture more
rapid. MT rotational diffusion and lateral capture were found to be
important mechanisms in fission yeast \cite{kalinina12}.  This work
suggested that MT rotational diffusion about centrosomes is important,
but because the model assumed fixed-length MTs it was unable to
directly evaluate the relative importance of rotational diffusion and
dynamic instability to KC capture.

We developed a biophysical model of KC capture in fission yeast that
includes MT dynamic instability and rotational diffusion, KC
diffusion, and interactions with the nuclear envelope
(fig.~\ref{fig:kc_capture}). We used the model to compare the time
course of KC capture in models with and without MT rotational
diffusion.  The mitotic MTs measured by Kalinina \textit{et al}.~were
primarily paused \cite{kalinina12}, and the measured dynamics appeared
slow compared to previous measurements \cite{sagolla03}. Therefore, we
performed further measurements of the dynamics of MTs in fission yeast
monopolar spindles with low-level fluorescent tubulin labeling. We
found more rapid dynamic instability of single MTs, with little
pausing (fig.~\ref{fig:experiment_data_kc}). This motivated us to
consider two dynamic instability models: a slow model with dynamics
measured by Kalinina \textit{et al}., and a fast model with dynamics
measured in this work (table \ref{table:kc_cap_params}).

We studied KC capture by single MTs in our model both for the
reference parameter sets and for broad ranges of parameters around the
reference sets. In all cases, KC capture occurred more quickly when
rotational diffusion about SPBs was added to a model with only dynamic
instability (fig.~\ref{fig:ttc_vs_l}): $\ttc$ decreased by 23\% on
average for the slow model, and by 9\% on average for the fast model,
similar to recent work \cite{cojoc16}. We found that the primary
determinant of the capture time was the mean MT length.  Longer MTs
have a larger surface on which the KC can bind, which suggests that
the capture time $\ttc \sim \meanl^{-1}$, as we observed.  Since the
MT growth speed and catastrophe frequency controlled the mean length
in the parameter regimes we studied, these parameters had the biggest
effect on the capture time (table \ref{table:sobol_indices},
fig.~\ref{fig:ttc_vs_vg_fcat_fixed}).  

For the experimentally measured mitotic polar MT lengths of $\sim 1$
\micron\ in fission yeast, adding rotational diffusion to our model
shortened the capture time by 14\% for the slow model
  and 6\% for the fast model.  This suggests that rotational
diffusion causes relatively small changes to the capture time
for typical conditions in fission yeast. We note that
  our model, like that of Kalinina et al.~\cite{kalinina12} considered
  the capturing MT only and neglected the bipolar spindle that
  assembles in prometaphase and the possibility of capturing MTs from
  the other spindle-pole body. Because the assembled spindle alters
  the rotational freedom of mitotic polar MTs, in future work it would
  be interesting to study how the presence of the spindle affects the
  kinetics of KC capture.
 
We further examined how MT length (fig.~\ref{fig:l_vs_params}) and
capture time (fig.~\ref{fig:ttc_vs_params}) varied with the growth
speed and catastrophe frequency. The mean length varied qualitatively
as expected from the predictions of analytic theory that neglects MT
interactions with the nuclear envelope. In the computational model,
changes from this theory became significant for longer MTs. Parameter
sets with higher growth speed and lower catastrophe frequency that
lead to longer mean MT lengths had shorter capture times.

The original search-and-capture model supposed that KC attachments to
MTs are end on, consistent with the results of
  Kalinina \textit{et al}.~\cite{kalinina12}, while others observed
primarily lateral attachments
\cite{rieder90,tanaka05,kitajima11,magidson11,magidson15}. Therefore,
we studied how the probability of lateral or end-on attachment varied
with MT length and the presence or absence of rotational diffusion in
the model (fig.~\ref{fig:lcap_vs_l}).  Lateral captures predominated,
even in the absence of MT rotational diffusion. Somewhat
counterintuitively, we found that adding rotational diffusion to the
model made end-on attachments more likely.  A diffusing MT searches a
larger volume of space than a rotationally constrained MT, an effect
which is largest for the MT tip. Therefore, rotational diffusion can
decrease lateral attachment.

Recently, Magidson et al.~proposed that initial lateral contacts are
important to the timing and low error rates of KC-spindle attachment
in human cells \cite{magidson15}. Although many proteins are known to
contribute to KC-MT attachment, including motors and non-motor
MT-binding proteins (the Ndc80 complex, other components of the KMN
network, Ska or Dam1, and others) \cite{cheeseman08}, the
contributions of end-on versus lateral KC-MT attachment pathways are
not fully understood.  Our work suggests that lateral captures are
enhanced when MT rotational diffusion about SPBs is decreased or
eliminated, as would be the case in large spindles with many long MTs.
Future work dissecting contributions of lateral and end-on attachment
mechanisms may contribute interesting additional insights into this
biologically important problem.

\section*{Author contributions}

RB, OS-S, CE, ZRG, AD, JRM, MAG, and MDB designed research; RB, OS-S,
CE, ZRG, PJF, SM, AC, and MDB performed research; RB, OS-S, CE, ZRG,
PJF, AD, MAG, and MDB contributed analytic tools; RB, OS-S, CE, ZRG,
PJF, SM, AC, and MDB analyzed data; RB, CE, ZRG, PJF, and MDB wrote
the manuscript.

\section*{Acknowledgements}

We thank Iain Hagan and Jonathan Millar for providing fission yeast
strains, Keith Gull for providing the TAT-1 antibody, and Nenad Pavin
for useful discussions.  This work was supported by NSF grants
DMR-0847685 and DMR-1551095 to MDB, MRSEC DMR-0820579 and DMR-1420736
to MAG, and CMMI-1454601 to AD; and NIH grants K25 GM110486 to MDB and
R01 GM033787 to JRM.  This work utilized the Janus supercomputer,
which is supported by the National Science Foundation (CNS-0821794),
the University of Colorado Boulder, the University of Colorado Denver,
and the National Center for Atmospheric Research. The Janus
supercomputer is operated by the University of Colorado Boulder.

\bibliography{zoterolibrary,bibliography}

\begin{thebibliography}{62}
\providecommand{\url}[1]{\texttt{#1}}
\providecommand{\urlprefix}{ }

\bibitem[McIntosh et~al.(2012)McIntosh, Molodtsov, and
  Ataullakhanov]{mcintosh12}
McIntosh, J.~R., M.~I. Molodtsov, and F.~I. Ataullakhanov, 2012.
\newblock Biophysics of mitosis.
\newblock \emph{Quarterly Reviews of Biophysics} 45:147--207.

\bibitem[Cottingham and Hoyt(1997)]{cottingham97}
Cottingham, F.~R., and M.~A. Hoyt, 1997.
\newblock Mitotic spindle positioning in {{Saccharomyces}} cerevisiae is
  accomplished by antagonistically acting microtubule motor proteins.
\newblock \emph{The Journal of Cell Biology} 138:1041--1053.

\bibitem[Goshima and Vale(2003)]{goshima03}
Goshima, G., and R.~D. Vale, 2003.
\newblock The roles of microtubule-based motor proteins in mitosis.
\newblock \emph{The Journal of Cell Biology} 162:1003--1016.

\bibitem[Grishchuk and McIntosh(2006)]{grishchuk06}
Grishchuk, E.~L., and J.~R. McIntosh, 2006.
\newblock Microtubule depolymerization can drive poleward chromosome motion in
  fission yeast.
\newblock \emph{The EMBO Journal} 25:4888--4896.

\bibitem[Schroer(2001)]{schroer01}
Schroer, T.~A., 2001.
\newblock Microtubules don and doff their caps: dynamic attachments at plus and
  minus ends.
\newblock \emph{Current Opinion in Cell Biology} 13:92--96.

\bibitem[Garcia et~al.(2002)Garcia, Koonrugsa, and Toda]{garcia02a}
Garcia, M.~A., N.~Koonrugsa, and T.~Toda, 2002.
\newblock Spindle\textendash{}kinetochore attachment requires the combined
  action of {{Kin I}}-like {{Klp5}}/6 and {{Alp14}}/{{Dis1-MAPs}} in fission
  yeast.
\newblock \emph{The EMBO Journal} 21:6015.

\bibitem[Akhmanova and Steinmetz(2008)]{akhmanova08}
Akhmanova, A., and M.~O. Steinmetz, 2008.
\newblock Tracking the ends: a dynamic protein network controls the fate of
  microtubule tips.
\newblock \emph{Nature Reviews Molecular Cell Biology} 9:309--322.

\bibitem[Duesberg et~al.(2006)Duesberg, Li, Fabarius, and Hehlmann]{duesberg06}
Duesberg, P., R.~Li, A.~Fabarius, and R.~Hehlmann, 2006.
\newblock Aneuploidy and cancer: from correlation to causation.
\newblock \emph{In} T.~Dittmar, K.~S. Zaenker, and A.~Schmidt, editors,
  Infection and {{Inflammation}}: {{Impacts}} on {{Oncogenesis}}, {Karger,
  Basel}, volume~13 of \emph{Contrib Microbiol}, 16--44.

\bibitem[Mitchison and Kirschner(1984)]{mitchison84}
Mitchison, T., and M.~Kirschner, 1984.
\newblock Dynamic instability of microtubule growth.
\newblock \emph{Nature} 312:237--242.

\bibitem[Mitchison and Kirschner(1985)]{mitchison85}
Mitchison, T.~J., and M.~W. Kirschner, 1985.
\newblock Properties of the kinetochore in vitro. {{II}}. {{Microtubule}}
  capture and {{ATP}}-dependent translocation.
\newblock \emph{The Journal of Cell Biology} 101:766--777.

\bibitem[Hill(1985)]{hill85}
Hill, T.~L., 1985.
\newblock Theoretical problems related to the attachment of microtubules to
  kinetochores.
\newblock \emph{Proceedings of the National Academy of Sciences} 82:4404--4408.

\bibitem[Holy and Leibler(1994)]{holy94}
Holy, T.~E., and S.~Leibler, 1994.
\newblock Dynamic instability of microtubules as an efficient way to search in
  space.
\newblock \emph{Proceedings of the National Academy of Sciences} 91:5682--5685.

\bibitem[Heald and Khodjakov(2015)]{heald15}
Heald, R., and A.~Khodjakov, 2015.
\newblock Thirty years of search and capture: {{The}} complex simplicity of
  mitotic spindle assembly.
\newblock \emph{The Journal of Cell Biology} 211:1103--1111.

\bibitem[Rieder and Alexander(1990)]{rieder90}
Rieder, C.~L., and S.~P. Alexander, 1990.
\newblock Kinetochores are transported poleward along a single astral
  microtubule during chromosome attachment to the spindle in newt lung cells.
\newblock \emph{The Journal of Cell Biology} 110:81--95.

\bibitem[Paul et~al.(2009)Paul, Wollman, Silkworth, Nardi, Cimini, and
  Mogilner]{paul09}
Paul, R., R.~Wollman, W.~T. Silkworth, I.~K. Nardi, D.~Cimini, and A.~Mogilner,
  2009.
\newblock Computer simulations predict that chromosome movements and rotations
  accelerate mitotic spindle assembly without compromising accuracy.
\newblock \emph{Proceedings of the National Academy of Sciences}
  106:15708--15713.

\bibitem[Carazo-Salas et~al.(1999)Carazo-Salas, Guarguaglini, Gruss, Segref,
  Karsenti, and Mattaj]{carazo-salas99}
Carazo-Salas, R.~E., G.~Guarguaglini, O.~J. Gruss, A.~Segref, E.~Karsenti, and
  I.~W. Mattaj, 1999.
\newblock Generation of {{GTP}}-bound {{Ran}} by {{RCC1}} is required for
  chromatin-induced mitotic spindle formation.
\newblock \emph{Nature} 400:178--181.

\bibitem[Wollman et~al.(2005)Wollman, Cytrynbaum, Jones, Meyer, Scholey, and
  Mogilner]{wollman05}
Wollman, R., E.~Cytrynbaum, J.~Jones, T.~Meyer, J.~Scholey, and A.~Mogilner,
  2005.
\newblock Efficient {{Chromosome Capture Requires}} a {{Bias}} in the
  `{{Search}}-and-{{Capture}}' {{Process}} during {{Mitotic-Spindle Assembly}}.
\newblock \emph{Current Biology} 15:828--832.

\bibitem[O'Connell et~al.(2009)O'Connell, Lon{\v c}arek, Kal{\'a}b, and
  Khodjakov]{oconnell09}
O'Connell, C.~B., J.~Lon{\v c}arek, P.~Kal{\'a}b, and A.~Khodjakov, 2009.
\newblock Relative contributions of chromatin and kinetochores to mitotic
  spindle assembly.
\newblock \emph{The Journal of Cell Biology} 187:43--51.

\bibitem[Magidson et~al.(2011)Magidson, O'Connell, Lon{\v c}arek, Paul,
  Mogilner, and Khodjakov]{magidson11}
Magidson, V., C.~B. O'Connell, J.~Lon{\v c}arek, R.~Paul, A.~Mogilner, and
  A.~Khodjakov, 2011.
\newblock The {{Spatial Arrangement}} of {{Chromosomes}} during {{Prometaphase
  Facilitates Spindle Assembly}}.
\newblock \emph{Cell} 146:555--567.

\bibitem[Magidson et~al.(2015)Magidson, Paul, Yang, Ault, O'Connell,
  Tikhonenko, McEwen, Mogilner, and Khodjakov]{magidson15}
Magidson, V., R.~Paul, N.~Yang, J.~G. Ault, C.~B. O'Connell, I.~Tikhonenko,
  B.~F. McEwen, A.~Mogilner, and A.~Khodjakov, 2015.
\newblock Adaptive changes in the kinetochore architecture facilitate proper
  spindle assembly.
\newblock \emph{Nature Cell Biology} .

\bibitem[Witt et~al.(1980)Witt, Ris, and Borisy]{witt80}
Witt, P.~L., H.~Ris, and G.~G. Borisy, 1980.
\newblock Origin of kinetochore microtubules in {{Chinese}} hamster ovary
  cells.
\newblock \emph{Chromosoma} 81:483--505.

\bibitem[Kitamura et~al.(2010)Kitamura, Tanaka, Komoto, Kitamura, Antony, and
  Tanaka]{kitamura10}
Kitamura, E., K.~Tanaka, S.~Komoto, Y.~Kitamura, C.~Antony, and T.~U. Tanaka,
  2010.
\newblock Kinetochores {{Generate Microtubules}} with {{Distal Plus Ends}}:
  {{Their Roles}} and {{Limited Lifetime}} in {{Mitosis}}.
\newblock \emph{Developmental Cell} 18:248--259.

\bibitem[Tanaka et~al.(2005)Tanaka, Mukae, Dewar, {van Breugel}, James,
  Prescott, Antony, and Tanaka]{tanaka05}
Tanaka, K., N.~Mukae, H.~Dewar, M.~{van Breugel}, E.~K. James, A.~R. Prescott,
  C.~Antony, and T.~U. Tanaka, 2005.
\newblock Molecular mechanisms of kinetochore capture by spindle microtubules.
\newblock \emph{Nature} 434:987--994.

\bibitem[Kalinina et~al.(2012)Kalinina, Nandi, Delivani, Chac{\'o}n, Klemm,
  Ramunno-Johnson, Krull, Lindner, Pavin, and Toli{\'c}-Norrelykke]{kalinina12}
Kalinina, I., A.~Nandi, P.~Delivani, M.~R. Chac{\'o}n, A.~H. Klemm,
  D.~Ramunno-Johnson, A.~Krull, B.~Lindner, N.~Pavin, and I.~M.
  Toli{\'c}-Norrelykke, 2012.
\newblock Pivoting of microtubules around the spindle pole accelerates
  kinetochore capture.
\newblock \emph{Nature Cell Biology} .

\bibitem[Kitajima et~al.(2011)Kitajima, Ohsugi, and Ellenberg]{kitajima11}
Kitajima, T.~S., M.~Ohsugi, and J.~Ellenberg, 2011.
\newblock Complete {{Kinetochore Tracking Reveals Error-Prone Homologous
  Chromosome Biorientation}} in {{Mammalian Oocytes}}.
\newblock \emph{Cell} 146:568--581.

\bibitem[Gopalakrishnan and Govindan(2011)]{gopalakrishnan11}
Gopalakrishnan, M., and B.~S. Govindan, 2011.
\newblock A {{First-Passage-Time Theory}} for {{Search}} and {{Capture}} of
  {{Chromosomes}} by {{Microtubules}} in {{Mitosis}}.
\newblock \emph{Bulletin of Mathematical Biology} 73:2483--2506.

\bibitem[Cojoc et~al.(2016)Cojoc, Florescu, Krull, Klemm, Pavin, J{\"u}licher,
  and Toli{\'c}]{cojoc16}
Cojoc, G., A.-M. Florescu, A.~Krull, A.~H. Klemm, N.~Pavin, F.~J{\"u}licher,
  and I.~M. Toli{\'c}, 2016.
\newblock Paired arrangement of kinetochores together with microtubule pivoting
  and dynamics drive kinetochore capture in meiosis {{I}}.
\newblock \emph{Scientific Reports} 6:25736.

\bibitem[Gao et~al.(2015{\natexlab{a}})Gao, Blackwell, Glaser, Betterton, and
  Shelley]{gao15a}
Gao, T., R.~Blackwell, M.~A. Glaser, M.~D. Betterton, and M.~J. Shelley, 2015.
\newblock Multiscale modeling and simulation of microtubule-motor-protein
  assemblies.
\newblock \emph{Physical Review E} 92:062709.

\bibitem[Gao et~al.(2015{\natexlab{b}})Gao, Blackwell, Glaser, Betterton, and
  Shelley]{gao15}
Gao, T., R.~Blackwell, M.~A. Glaser, M.~Betterton, and M.~J. Shelley, 2015.
\newblock Multiscale {{Polar Theory}} of {{Microtubule}} and {{Motor-Protein
  Assemblies}}.
\newblock \emph{Physical Review Letters} 114:048101.

\bibitem[Kuan et~al.(2015)Kuan, Blackwell, Hough, Glaser, and
  Betterton]{kuan15}
Kuan, H.-S., R.~Blackwell, L.~E. Hough, M.~A. Glaser, and M.~D. Betterton,
  2015.
\newblock Hysteresis, reentrance, and glassy dynamics in systems of
  self-propelled rods.
\newblock \emph{Physical Review E} 92:060501.

\bibitem[Blackwell et~al.(2016)Blackwell, Sweezy-Schindler, Baldwin, Hough,
  Glaser, and Betterton]{blackwell16}
Blackwell, R., O.~Sweezy-Schindler, C.~Baldwin, L.~E. Hough, M.~A. Glaser, and
  M.~D. Betterton, 2016.
\newblock Microscopic origins of anisotropic active stress in motor-driven
  nematic liquid crystals.
\newblock \emph{Soft Matter} .

\bibitem[Sagolla et~al.(2003)Sagolla, Uzawa, and Cande]{sagolla03}
Sagolla, M.~J., S.~Uzawa, and W.~Z. Cande, 2003.
\newblock Individual microtubule dynamics contribute to the function of mitotic
  and cytoplasmic arrays in fission yeast.
\newblock \emph{Journal of Cell Science} 116:4891--4903.

\bibitem[Laan et~al.(2012)Laan, Pavin, Husson, Romet-Lemonne, van Duijn,
  L{\'o}pez, Vale, J{\"u}licher, Reck-Peterson, and Dogterom]{laan12}
Laan, L., N.~Pavin, J.~Husson, G.~Romet-Lemonne, M.~van Duijn, M.~P. L{\'o}pez,
  R.~D. Vale, F.~J{\"u}licher, S.~L. Reck-Peterson, and M.~Dogterom, 2012.
\newblock Cortical {{Dynein Controls Microtubule Dynamics}} to {{Generate
  Pulling Forces}} that {{Position Microtubule Asters}}.
\newblock \emph{Cell} 148:502--514.

\bibitem[Pavin et~al.(2012)Pavin, Laan, Ma, Dogterom, and
  J{\"u}licher]{pavin12}
Pavin, N., L.~Laan, R.~Ma, M.~Dogterom, and F.~J{\"u}licher, 2012.
\newblock Positioning of microtubule organizing centers by cortical pushing and
  pulling forces.
\newblock \emph{New Journal of Physics} 14:105025.

\bibitem[Ma et~al.(2014)Ma, Laan, Dogterom, Pavin, and J{\"u}licher]{ma14}
Ma, R., L.~Laan, M.~Dogterom, N.~Pavin, and F.~J{\"u}licher, 2014.
\newblock General theory for the mechanics of confined microtubule asters.
\newblock \emph{New Journal of Physics} 16:013018.

\bibitem[Tischer et~al.(2009)Tischer, Brunner, and Dogterom]{tischer09}
Tischer, C., D.~Brunner, and M.~Dogterom, 2009.
\newblock Force- and kinesin-8-dependent effects in the spatial regulation of
  fission yeast microtubule dynamics.
\newblock \emph{Molecular Systems Biology} 5:1--10.

\bibitem[Dogterom and Yurke(1997)]{dogterom97}
Dogterom, M., and B.~Yurke, 1997.
\newblock Measurement of the {{Force-Velocity Relation}} for {{Growing
  Microtubules}}.
\newblock \emph{Science} 278:856--860.

\bibitem[Janson et~al.(2003)Janson, {de Dood}, and Dogterom]{janson03}
Janson, M.~E., M.~E. {de Dood}, and M.~Dogterom, 2003.
\newblock Dynamic instability of microtubules is regulated by force.
\newblock \emph{The Journal of Cell Biology} 161:1029--1034.

\bibitem[Yamagishi et~al.(2012)Yamagishi, Yang, Tanno, and
  Watanabe]{yamagishi12}
Yamagishi, Y., C.~H. Yang, Y.~Tanno, and Y.~Watanabe, 2012.
\newblock {{MPS1}}/{{Mph1}} phosphorylates the kinetochore protein
  {{KNL1}}/{{Spc7}} to recruit {{SAC}} components.
\newblock \emph{Nature Cell Biology} 14:746--752.

\bibitem[Hagan and Yanagida(1990)]{hagan90}
Hagan, I., and M.~Yanagida, 1990.
\newblock Novel potential mitotic motor protein encoded by the fission yeast
  cut7+ gene.
\newblock \emph{Nature} 347:563--566.

\bibitem[Costa et~al.(2013)Costa, Fu, Syrovatkina, and Tran]{costa13}
Costa, J., C.~Fu, V.~Syrovatkina, and P.~T. Tran, 2013.
\newblock Chapter 24 - {{Imaging Individual Spindle Microtubule Dynamics}} in
  {{Fission Yeast}}.
\newblock \emph{In} J.~J.~C. Wilson, and Leslie, editors, Methods in {{Cell
  Biology}}, {Academic Press}, volume 115 of \emph{Microtubules, in Vitro},
  385--394.

\bibitem[Demchouk et~al.(2011)Demchouk, Gardner, and Odde]{demchouk11}
Demchouk, A.~O., M.~K. Gardner, and D.~J. Odde, 2011.
\newblock Microtubule {{Tip Tracking}} and {{Tip Structures}} at the
  {{Nanometer Scale Using Digital Fluorescence Microscopy}}.
\newblock \emph{Cellular and Molecular Bioengineering} 4:192--204.

\bibitem[Prahl et~al.(2014)Prahl, Castle, Gardner, and Odde]{prahl14}
Prahl, L.~S., B.~T. Castle, M.~K. Gardner, and D.~J. Odde, 2014.
\newblock Chapter {{Three}} - {{Quantitative Analysis}} of {{Microtubule
  Self}}-assembly {{Kinetics}} and {{Tip Structure}}.
\newblock \emph{In} R.~D. Vale, editor, Methods in {{Enzymology}}, {Academic
  Press}, volume 540 of \emph{Reconstituting the Cytoskeleton}, 35--52.

\bibitem[Ding et~al.(1993)Ding, McDonald, and McIntosh]{ding93}
Ding, R., K.~L. McDonald, and J.~R. McIntosh, 1993.
\newblock Three-dimensional reconstruction and analysis of mitotic spindles
  from the yeast, {{Schizosaccharomyces}} pombe.
\newblock \emph{The Journal of Cell Biology} 120:141--151.

\bibitem[Alberts et~al.(2008)Alberts, Johnson, Lewis, Raff, Roberts, and
  Walter]{alberts08}
Alberts, B., A.~Johnson, J.~Lewis, M.~Raff, K.~Roberts, and P.~Walter, 2008.
\newblock Molecular {{Biology}} of the {{Cell}}.
\newblock {Garland}, New York, 5th edition.

\bibitem[Snaith et~al.(2010)Snaith, Anders, Samejima, and Sawin]{snaith10}
Snaith, H.~A., A.~Anders, I.~Samejima, and K.~E. Sawin, 2010.
\newblock Chapter 9 - {{New}} and {{Old Reagents}} for {{Fluorescent Protein
  Tagging}} of {{Microtubules}} in {{Fission Yeast}}: {{Experimental}} and
  {{Critical Evaluation}}.
\newblock \emph{In} {Lynne Cassimeris and Phong Tran}, editor, Methods in
  {{Cell Biology}}, {Academic Press}, volume Volume 97, 147--172.

\bibitem[Bratman and Chang(2007)]{bratman07}
Bratman, S.~V., and F.~Chang, 2007.
\newblock Stabilization of {{Overlapping Microtubules}} by {{Fission Yeast
  CLASP}}.
\newblock \emph{Developmental Cell} 13:812--827.

\bibitem[Bratman and Chang(2008)]{bratman08}
Bratman, S.~V., and F.~Chang, 2008.
\newblock Mechanisms for maintaining microtubule bundles.
\newblock \emph{Trends in Cell Biology} 18:580--586.

\bibitem[Fygenson et~al.(1994)Fygenson, Braun, and Libchaber]{fygenson94}
Fygenson, D.~K., E.~Braun, and A.~Libchaber, 1994.
\newblock Phase diagram of microtubules.
\newblock \emph{Physical Review E} 50:1579.

\bibitem[Gardner et~al.(2008)Gardner, Bouck, Paliulis, Meehl, O'Toole, Haase,
  Soubry, Joglekar, Winey, Salmon, Bloom, and Odde]{gardner08}
Gardner, M.~K., D.~C. Bouck, L.~V. Paliulis, J.~B. Meehl, E.~T. O'Toole,
  J.~Haase, A.~Soubry, A.~P. Joglekar, M.~Winey, E.~D. Salmon, K.~Bloom, and
  D.~J. Odde, 2008.
\newblock Chromosome {{Congression}} by {{Kinesin}}-5 {{Motor-Mediated
  Disassembly}} of {{Longer Kinetochore Microtubules}}.
\newblock \emph{Cell} 135:894--906.

\bibitem[Dogterom and Leibler(1993)]{dogterom93}
Dogterom, M., and S.~Leibler, 1993.
\newblock Physical aspects of the growth and regulation of microtubule
  structures.
\newblock \emph{Physical Review Letters} 70:1347--1350.

\bibitem[Ghanem and Spanos(2002)]{Ghanem03}
Ghanem, R., and P.~Spanos, 2002.
\newblock Stochastic Finite Elements: A Spectral Approach.
\newblock Dover.

\bibitem[Xiu and Karniadakis(2002)]{Xiu02}
Xiu, D., and G.~Karniadakis, 2002.
\newblock The {W}iener-{A}skey polynomial chaos for stochastic differential
  equations.
\newblock \emph{SIAM Journal on Scientific Computing} 24:619--644.

\bibitem[Doostan and Owhadi(2011)]{doostan11a}
Doostan, A., and H.~Owhadi, 2011.
\newblock A non-adapted sparse approximation of {PDEs} with stochastic inputs.
\newblock \emph{Journal of Computational Physics} 230:3015--3034.

\bibitem[Hampton and Doostan(2015)]{hampton15a}
Hampton, J., and A.~Doostan, 2015.
\newblock Compressive sampling of polynomial chaos expansions: Convergence
  analysis and sampling strategies.
\newblock \emph{Journal of Computational Physics} 280:363 -- 386.

\bibitem[Hadigol et~al.(2015)Hadigol, Maute, and Doostan]{hadigol15}
Hadigol, M., K.~Maute, and A.~Doostan, 2015.
\newblock On uncertainty quantification of lithium-ion batteries:
  {{Application}} to an {{LiC6}}/{{LiCoO2}} cell.
\newblock \emph{Journal of Power Sources} 300:507--524.

\bibitem[Sobol'(1990)]{sobol90}
Sobol', I., 1990.
\newblock On sensitivity estimation for nonlinear mathematical models.
\newblock \emph{Matematicheskoe Modelirovanie} 2:112--118.

\bibitem[Sudret(2008)]{sudret08}
Sudret, B., 2008.
\newblock Global sensitivity analysis using polynomial chaos expansions.
\newblock \emph{Reliability Engineering and System Safety} 93:964 -- 979.

\bibitem[Cheeseman and Desai(2008)]{cheeseman08}
Cheeseman, I.~M., and A.~Desai, 2008.
\newblock Molecular architecture of the kinetochore\textendash{}microtubule
  interface.
\newblock \emph{Nature Reviews Molecular Cell Biology} 9:33--46.

\bibitem[Lowen(1994)]{lowen94}
Lowen, H., 1994.
\newblock Brownian dynamics of hard spherocylinders.
\newblock \emph{Physical Review E} 50:1232--1242.

\bibitem[Weeks et~al.(1971)Weeks, Chandler, and Andersen]{weeks71}
Weeks, J.~D., D.~Chandler, and H.~C. Andersen, 1971.
\newblock Role of {{Repulsive Forces}} in {{Determining}} the {{Equilibrium
  Structure}} of {{Simple Liquids}}.
\newblock \emph{The Journal of Chemical Physics} 54:5237--5247.

\bibitem[Moreno et~al.(1991)Moreno, Klar, and Nurse]{moreno91}
Moreno, S., A.~Klar, and P.~Nurse, 1991.
\newblock Molecular genetic analysis of fission yeast {{Schizosaccharomyces}}
  pombe.
\newblock \emph{In} Guide to {{Yeast Genetics}} and {{Molecular Biology}},
  {Academic Press}, volume Volume 194, 795--823.

\end{thebibliography}

\pagebreak
\setcounter{equation}{0}
\setcounter{figure}{0}
\setcounter{table}{0}
\setcounter{page}{1}
\renewcommand\thefigure{S\arabic{figure}}    
\renewcommand\theequation{S\arabic{equation}}    
\renewcommand\thetable{S\arabic{table}}   

\newcommand{\beq}{\begin{equation}}
\newcommand{\eeq}{\end{equation}}
\newcommand{\bD}{{\bf D}}
\newcommand{\bI}{{\bf I}}
\newcommand{\bQ}{{\bf Q}}
\newcommand{\bX}{{\bf X}}
\newcommand{\bF}{{\bf F}}
\newcommand{\bU}{{\bf U}}
\newcommand{\bSigma}{{\bf \Sigma}}
\newcommand{\bk}{{\bf k}}
\newcommand{\bu}{{\bf u}}
\newcommand{\br}{{\bf r}}
\newcommand{\bx}{{\bf x}}
\newcommand{\bp}{{\bf p}}
\newcommand{\bn}{{\bf n}}
\newcommand{\bq}{{\bf q}}
\newcommand{\Fs}{F_s}

\begin{center}
\textbf{\large Supporting Material}
\end{center}

\section{Model}

MTs are rigid spherocylinders (cylinders with hemispherical ends) with
length $L(t)$ and diameter $\sigma_{MT}$. One end of the MT is fixed
to a point on the nuclear envelope.  The equations of motion for
microtubule reorientation are
\begin{equation}
  \label{eq:brownianrot}
  {\bu}_i(t + \delta t) = {\bu}_i(t) + {D_\theta(L) \over {k_B T}} {\mathbf
    T}_i(t) \times {\bu}_i(t) \delta t + \delta {\bu}_i(t),
\end{equation}
where $D_\theta(L)$ is the rotational diffusion coefficient,
${\mathbf T}_i(t)$ is the systematic torque on particle $i$, and the
random reorientation $\delta {\bu}_i(t)$ is Gaussian-distributed, with
variance
$ \left\langle \delta {\bu}_i(t) \delta {\bu}_i(t) \right\rangle = 2
D_\theta(L) \left[ {\mathbf I} - {\bu}_i(t) {\bu}_i(t) \right] \delta
t$.
MTs have a length-dependent rotational diffusion coefficient
$D_{\theta}\left(L\right) \sim L^{-3}$. Using the formula for
spherocylinder rotational diffusion from L\"{o}wen \textit{et al}.
\cite{lowen94}, we calculated $D_{\theta}\left(L\right)$ for each
timestep of the simulation:
\begin{equation}
\label{eq:mt_rot_diff}
D_{\theta}\left(L(t)\right) = \frac{3k_{b}T}{\pi\eta(L(t)+1)^{3}}(\ln a_{MT} - 0.662 + 0.917/a_{MT} - 0.050/a_{MT}^{2}),
\end{equation}
where $\eta$ is the fluid viscosity, and $a_{MT}$ =
$L(t)/\sigma_{MT}$.

The MT minimum length is 4 $\sigma_{MT}$.  We tested shorter minimum
lengths, but found that decreasing below 4 $\sigma_{MT}$ did not
significantly change the capture time; this choice of minimum length
makes the simulations more stable.

KCs are spheres with diameter $\sigma_{KC}$ and diffusion coefficient
$D_{kc}$ (Table \ref{table:kc_cap_params}) that obey the equation of
motion
\begin{equation}
  \label{eq:browniansphere}
  {\br}_i (t + \delta t) = {\br}_i (t) + {D_{kc} \over {k_B T}} {\mathbf
  F}_{i}(t) + \delta {\br}_i (t),
\end{equation}
with Gaussian random displacements with variance
$ \langle \delta {\br}_{i}(t) \delta {\br}_{i}(t) \rangle =
2D_{kc}\delta t$.

\begin{figure}
  \centering
  \includegraphics[width=5 cm]{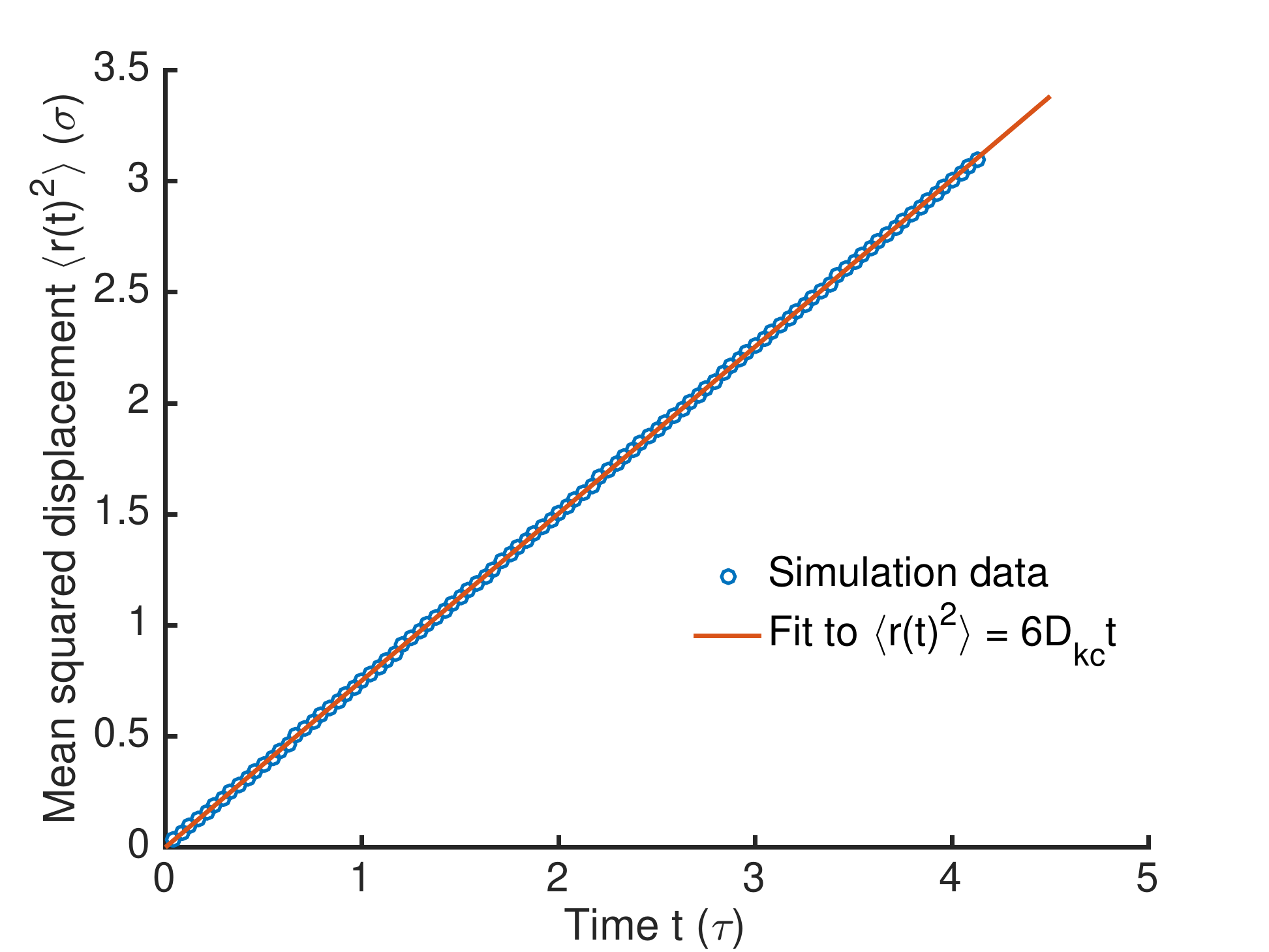}
  \includegraphics[width=5 cm]{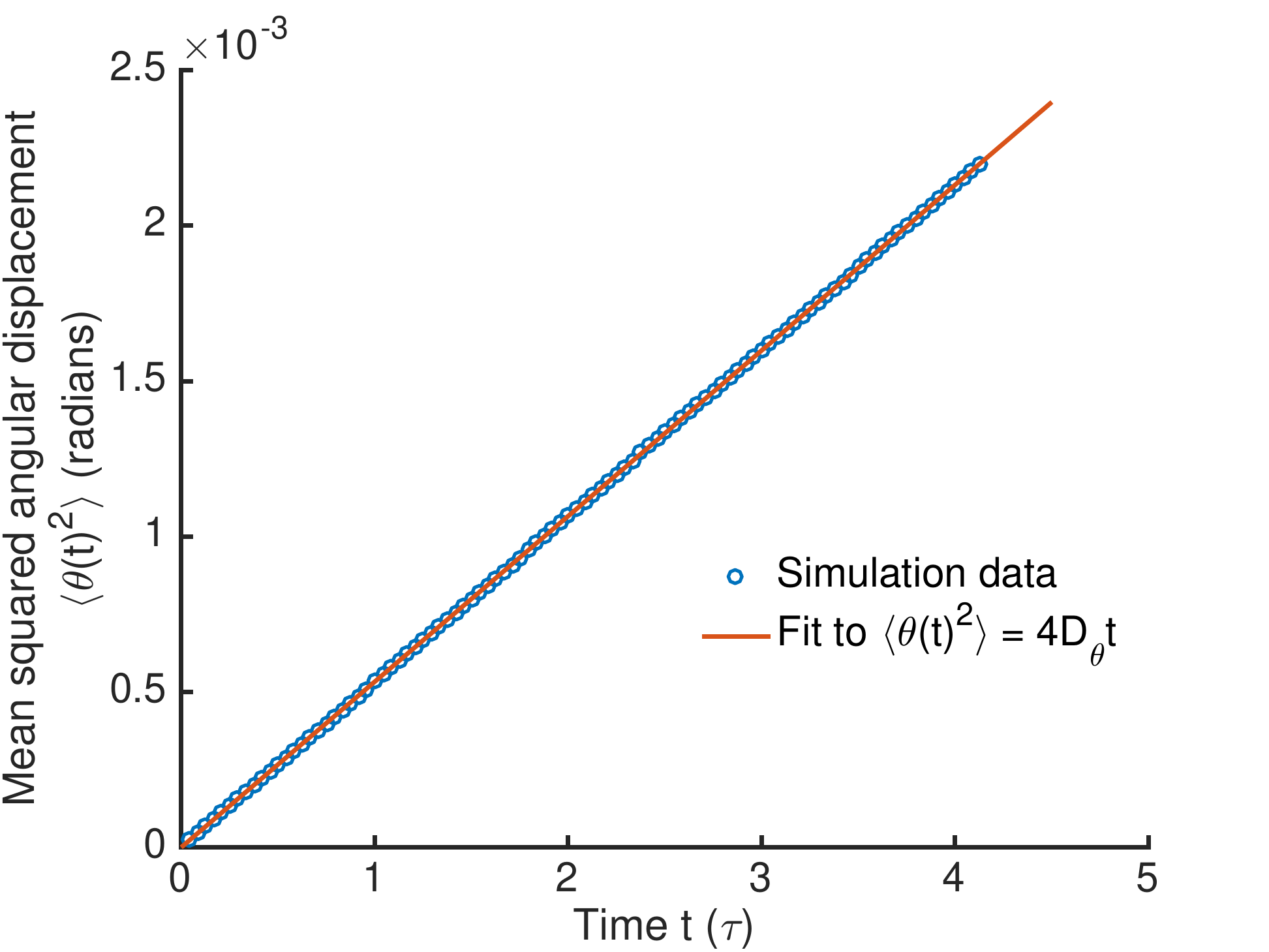}
  \includegraphics[width=5 cm]{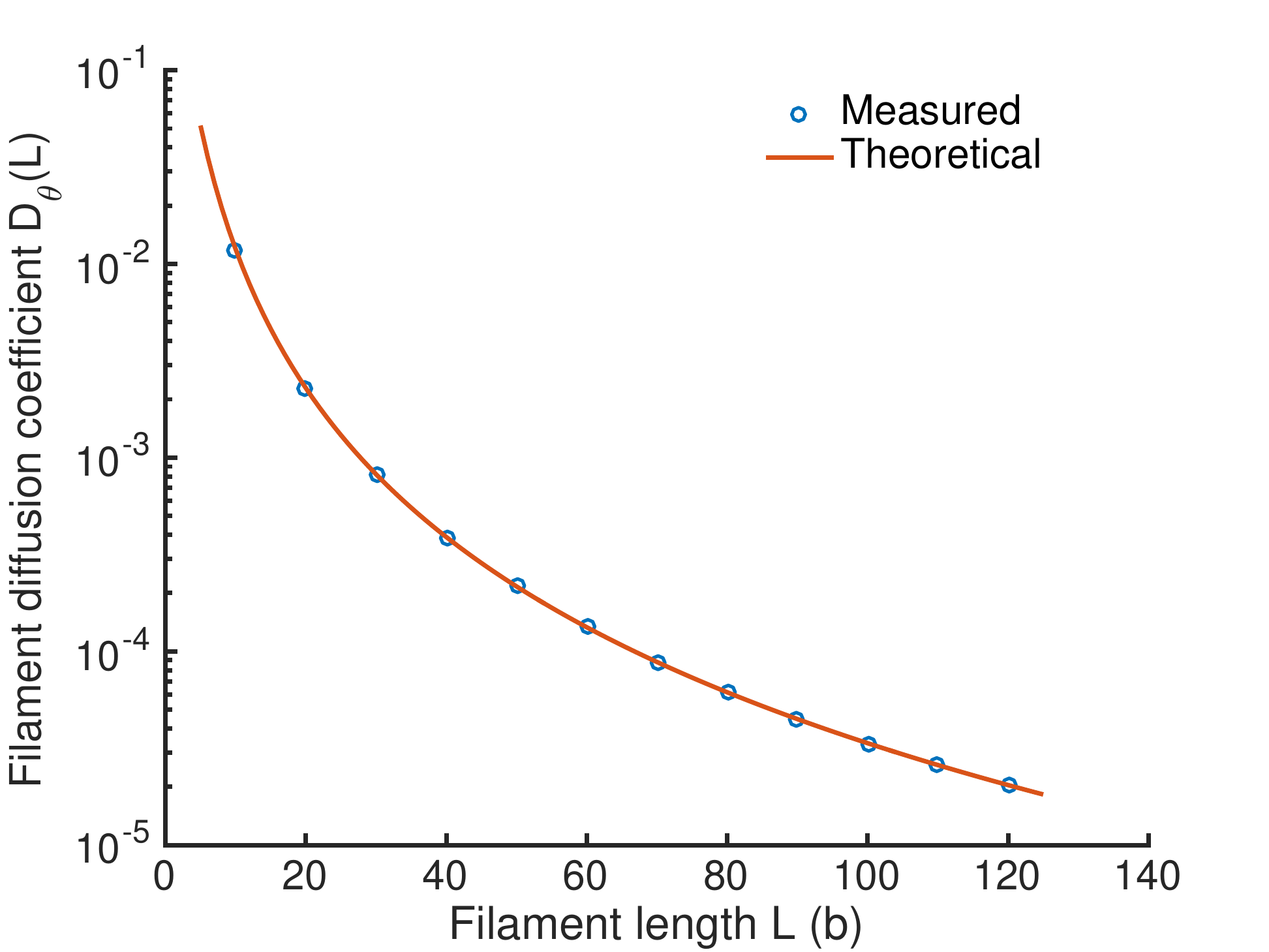}
  \caption {\footnotesize Diffusion coefficient verification.  (a)
    Kinetochore mean-squared displacement versus time for a 0.2 $\mu$m
    diameter kinetochore. Points indicate measured values from a
    simulation with no boundary; The line is a linear fit of
    $\langle r(t)^2\rangle = 6D_{\textrm{kc}}t$.  (b) Filament
    mean-squared angular displacement versus time. Points indicate
    measured values from a simulation with no boundary. The line is a
    linear fit of $\langle\theta(t)^2\rangle = 4D_{\theta}t$.  (c)
    Rotational diffusion coefficients versus microtubule length for
    filaments of fixed length. Points are measured values from a
    simulation with no boundary.  The line is the diffusion
    coefficient from L\"owen et al.~\cite{lowen94}.  Values in all
    plots are in dimensionless simulation units.  }
  \label{suppfig:diff_verification}
\end{figure}

To account for variations in the relative diffusion of KCs and MTs, we
defined the MT and KC diffusion coefficients independently
(fig.~\ref{suppfig:diff_verification}).  We nondimensionalized the
parameters using the reference length $\sigma_{MT}$ of a MT diameter
(25 nm), and the reference diffusion coefficient $D_0$; 
these together determine the unit of time
$\tau = \sigma_{MT}^2 / D_0 $.  A kinetochore of size 200nm
has a diffusion constant of $D_{kc} = 5.9 \times 10^{-4} $ \micron$^{2}$ \pers.
Using Stokes calculation for D, we see that $D_{kc} = \frac{1}{8} D_{0}$, yielding
$D_{0} = 4.72 \times 10^{-3} $ \micron$^{2}$ \pers.

\subsection{Boundary interactions}

To ensure that MTs and KCs remained within the nuclear envelope, both
MT free ends and KCs interacted with the envelope via the
Weeks-Chandler-Anderson potential \cite{weeks71}
\begin{equation}
  u_{\rm wca,fil}(r_{\rm min}) = \left\{
    \begin{array}{ll}
      4 k_B T \left[ \left( {\sigma \over r_{\rm min}} \right)^{12} -
      \left( {\sigma \over r_{\rm min}} \right)^6 \right] + k_BT, &
     r_{\rm min} < 2^{1/6} \sigma \\
      0, & r_{\rm min} \geq 2^{1/6} \sigma,
    \end{array}
  \right.
\end{equation}
where $r_{\textrm{min}}$ is minimum distance between the free end of
the filament and the enclosing sphere with radius $R + {\sigma / 2}$,
and $\sigma$ is the finite distance at which the potential goes to
zero.  This allows for smooth continuation of the dynamics at the
boundary; the nuclear envelope then has an effective radius of $R$.

Kinetochores have a similar interaction with the envelope 
\begin{equation}
  u_{\rm wca,kc}(r_{\rm min}) = \left\{
    \begin{array}{ll}
      4 k_B T \left[ \left( {\sigma_{\rm kc}/2 + \sigma/2 \over r_{\rm min}} \right)^{12} -
      \left( {\sigma_{\rm kc}/ + \sigma/2 \over r_{\rm min}} \right)^6 \right] + k_BT, &
     r_{\rm min} < 2^{1/6} (\sigma_{\rm kc} + \sigma)/2\\
      0, & r_{\rm min} \geq 2^{1/6} (\sigma_{\rm kc} + \sigma)/2,
    \end{array}
  \right.
\end{equation}
where again an enclosing boundary is a sphere with radius $R + {\sigma / 2}$.

\subsection{Dynamic instability models}

We consider two dynamic instability models.  The first is the
original two-state dynamic instability model of Dogterom and Leibler
\cite{dogterom93}.  The microtubule (MT) grows (shrinks) at a constant
velocity $v_{g}$ ($v_{s}$), with transition frequencies
$f_{c}$ ($f_{r}$) to the other state. This leads to an exponential
distribution of MT lengths in the bounded growth regime, with mean
length
\begin{equation}
\label{eq:two_state_di_meanl}
\meanl = \frac{v_g v_s}{v_s f_c - v_g f_r}.
\end{equation}
If the second term in the denominator is small, which commonly occurs
in cells when the rescue frequency is low, we have $v_g f_r \ll v_s
f_c$ and this becomes approximately
\begin{equation}
\label{eq:two_meanl_approx}
\meanl \approx \frac{v_g }{f_c }.
\end{equation}

\subsubsection{Slow dynamic instability}

We consider MTs that can be in 3 states: growing, shrinking, or
pausing. In the paused state, the MT length doesn't change with
time. As in previous work, we consider the probability distribution of
MT state and length $P_s(z,t)$, where $s$ labels the state ($+$, $-$,
and $0$ label growing, shrinking, and pausing states), $z$ length, and
$t$ time. The equations of the three-state model are
\begin{eqnarray}
\label{eq:pp_prob_density}
\frac{\partial P_{+}(z,t)}{\partial t}  &=& -f_{+0} P_{+} + f_{0+}
                                           P_{0} - v_{g}
                                           \frac{ \partial P_{+} }{ \partial z} ,\\ 
\label{eq:pm_prob_density}
\frac{\partial  P_{-}(z,t)}{\partial t} &=& -f_{-0} P_{-} + f_{0-}
                                           P_{0} + v_{s}
                                           \frac{ \partial P_{-}}{ \partial
                                           z} , \\
\label{eq:p0_prob_density}
\frac{\partial P_{0}(z,t)}{\partial t}  &=& -(f_{0-} + f_{0+}) P_{0} +
                                           f_{+0} P_{+} + f_{-0}
                                           P_{-}. 
\end{eqnarray}
Here $f_{+ 0 }$ is the frequency of transitions from the growing state
to the paused state, and so on. To determine the mean length
$\langle L \rangle$, we study the steady state equations. After
summing the steady state equations, the state switching terms cancel
and we have
\begin{equation}
v_{+} \frac{ \partial P_{+}}{ \partial z}  = v_{-} \frac{ \partial P_{-}
}{ \partial z}.  
\end{equation}
Integrating both sides of the equation with respect to $z$ and setting
the integration constant to 0 (as required for a bounded length
distribution for which the probability density goes to zero for large
$z$) gives 
\begin{equation}
P_{-} = \frac{v_{+}}{v_{-}} P_{+}.
\end{equation}
Similarly, the steady-state of equation \eqref{eq:p0_prob_density}
gives
\begin{equation}
  P_{0} = \left( \frac{ f_{+0}+v_{+}f_{-0}/v_{-} }{ f_{0+} +
      f_{0-} } \right) P_{+}.
\end{equation}
Then equation \eqref{eq:pp_prob_density}  becomes
\begin{equation}
\frac{ \partial P_{+} }{ \partial z } =  \left(
  \frac{-f_{+0}}{v_{+}} + \frac{ f_{0+}f_{+0} }{v_{+} \left( f_{0+} +
      f_{0-} \right)} + \frac{ f_{0+}f_{-0} }{v_{-} \left( f_{0+} +
      f_{0-} \right)}\right) P_{+}. 
\end{equation}
If the factor in parentheses on the right side of this equation is
negative, then the length distribution is a exponential with mean
length 
\begin{equation}
\meanl = \left(
  \frac{f_{+0}}{v_{+}} - \frac{ f_{0+}f_{+0} }{v_{+} \left( f_{0+} +
      f_{0-} \right)} - \frac{ f_{0+}f_{-0} }{v_{-} \left( f_{0+} +
      f_{0-} \right)}\right)^{-1}.
  \label{eq:meanl-long}
\end{equation}

We consider the case in which transitions from pausing to growing do
not occur, so that $f_{0+} = 0$.  Then
\begin{equation}
  \meanl =   \frac{v_{g}}{f_{+0}} .
  \label{eq:meanl}
\end{equation}
Note the similarity of this expression to equation
\eqref{eq:two_meanl_approx} with $f_{+0}$ the effective catastrophe
frequency of the model.

\begin{figure}
  \centering
  \includegraphics[width=5 cm]{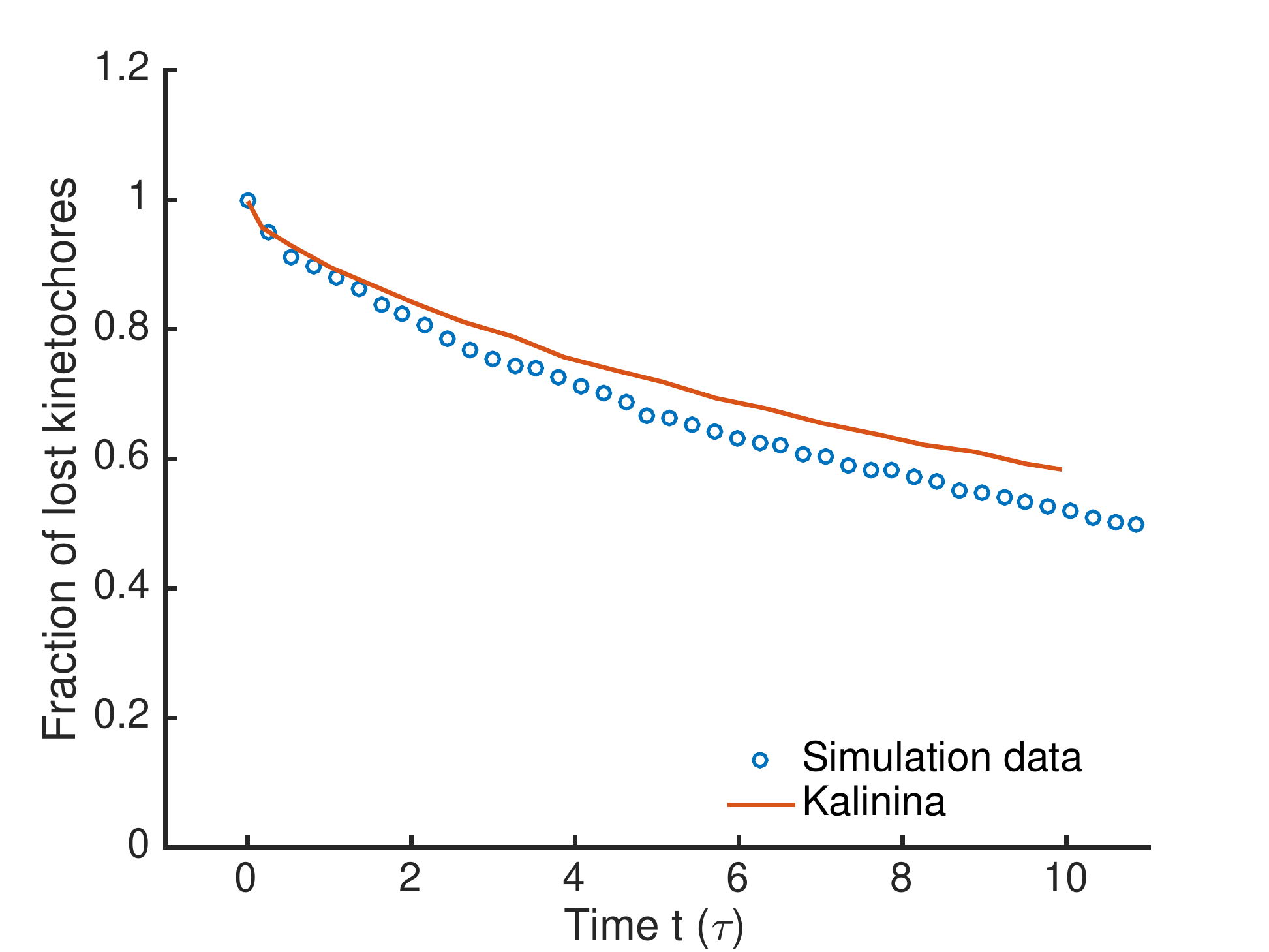}
  \caption {\footnotesize Comparison of fraction of lost kinetochores
    versus time in our model to those of Kalinina et
    al.~\cite{kalinina12}. For this comparison, we inserted
    kinetochores in the the nucleus uniformly 1.2 $\mu m$ from the
    base of the filament to mimic the approach of Kalinina et
    al. Microtubules were inserted with uniform angles that did not
    cause overlap with the boundary. The difference between the two
    models is likely due to the overlap criteria: Kalinina et al used
    an approximate angular criterion to detect overlaps that did not
    account for the non-zero volume of the filament
    \cite{kalinina12}. This effectively decreased the capture radius
    compared to our model.}
  \label{suppfig:kalinina_verification}
\end{figure}

\section{Experiment}

\subsection{Cut7-ts strain construction}

Cells were cultured using standard fission yeast techniques
\cite{moreno91}.  To construct \textit{cut7-ts} strains with
mCherry-tagged tubulin, parent strains with the low-expression
\textit{mCherry-atb2} MT marker \cite{yamagishi12} and the
\textit{cut7-24} allele \cite{hagan90} were crossed on malt extract
agar plates.  Zygotic asci were isolated by digestion with glusulase
(PerkinElmer, Waltham, Massachusetts), counted on a haemocytometer,
plated onto YE5S agar plates and allowed to grow into colonies.  The
colonies were replica plated sequentially onto two different plates,
first onto YEP (YE5S plus phloxin B) agar plates.  Colonies were
allowed to grow at 25\degree\ overnight and then placed at 36\degree.
After 1-2 days, dark pink colonies indicated dead or dying cells and
were selected from the parent plates as being positive for
\textit{cut7-24}.  Next, colonies were replica plated onto YES + 100
$\mu$g/mL nourseothricin (Gold Biotechnology, Olivette, Missouri) agar
plates.  Candidate colonies that grew were selected as positive for
\textit{mCherry-atb2}.  Candidates with both \textit{cut7-24} and
\textit{mCherry-atb2} were validated by imaging fluorescent MTs in
monopolar spindles at 36$^o$C using fluorescence microscopy on an
Axioplan II light microscope (Carl Zeiss, Jena, Germany) with a 100x,
1.45 NA Plan Fluor oil-immersion objective, a Bioptechs objective
heater (Bioptechs, Butler, PA) and a Photometrics Cascade 650 CCD
camera (Roper Scientific, Sarasota, Florida).

\subsection{Measurement of labeled tubulin fraction}

A preculture was grown in YPD from which a 50 mL YPD overnight cell
culture was grown to late log phase.  These cell cultures were
pelleted in a tabletop Beckman CS-6 centrifuge, the supernatant was
removed, and they were resuspended in 1 mL of NaCl (150 mM)-Tris (50
mM) buffer at pH 8.  The cells were washed twice in buffer and
resuspended in the same buffer plus 1/2 tablet of complete Mini
EDTA-free protease inhibitors (Roche Diagnostics, Mannheim, Germany).
The cells were then lysed with a FastPrep FP120 ribolyser (MP
Biomedicals, Santa Ana, CA).  Lysing tubes were prepared with 1 cm 0.5
$\mu$m glass beads on the bottom.  1 mL of cell suspension per tube
was lysed on setting 6 for 20 sec for a total of three runs.  In
between each run, the cells were placed on ice for 10 min.  The tubes
were spun in a tabletop centrifuge at 5,000 rpm for 6 min and then at
14,000 rpm for 30 min.  The clarified lysate was removed and flash
frozen with liquid nitrogen.

The cell lysate was mixed 1:1 with Laemmli sample buffer (Bio-Rad,
Hercules, CA) plus 5\% $\beta$-mercaptoethanol.  20 $\mu$L samples for
ten different lanes were prepared with serial dilutions ranging from
100\%-10\% of the original cell suspension concentration, and the
samples were boiled in water for 7 min.  The samples were run on Any
KD pre-cast gels in a BioRad Miniprotean II system (Bio-Rad, Hercules,
CA) with a Tris (25 mM), glycine (192 mM), 0.1\% SDS running buffer.
The gels were run for 90 min at 100V with ice packs around the gel
box.  

Immunoblots were then prepared using a modification of standard
techniques.  Once transfer sandwiches were assembled, samples were
transferred to either a nitrocellulose or a PVDF membrane in transfer
buffer consisting of Tris (25 mM), glycine (192 mM), 20\% methanol,
and 0.05\% SDS.  Blots were run for 1 hr at 80V.  The membranes were
transferred to a 5\% nonfat milk blocking solution in 1x TBS-T (Tris
Buffered Saline plus 0.05\% Tween 20) and blocked overnight.  The
blocked membrane was transferred to a sealable bag containing the
primary antibody TAT-1 at a 1:1000 dilution for 1 hr at room
temperature.  The blots were washed 3 times for 15 min in TBS-T and
transferred to a sealable bag containing the secondary antibody
Goat-anti-Mouse-AP (Bio-Rad, Hercules, CA) at a 1:1000 dilution for 1
hr at room temperature.  The blots were washed 3 times for 15 min in
TBS-T and transferred to 20 mL of 1-Step NBI/BCIP developing solution
(Thermo Scientific, Waltham, MA) for 30 min.


\begin{table*}[t]
  \centering
  \begin{tabular}{ lll }
    \textbf{Strain} & \textbf{Genotype} & \textbf{Source} \\
\hline
    Mcl 730 & nda3-KM311, cen2::kanr-ura4$^{+-}$lacOp his7$^+$::lacI-GFP, & This study \\
        & nmt1-GFP-pcp1$^+$::kanr, mcherry-atb2:natMX6, leu1-32, ura4-D18, h$^-$ & \\
    McI 804 & z:adh15:mcherry-atb2:natMX6, cut7-24, leu1-32, ura4-d18, h$^{90}$ & This study \\
    \textbf{Original Strains} & & \\
    McI 728 & z:adh15:mcherry-atb2:natMX6, leu1-32, ura4-d18, h$^{+}$ & Y. Wantanabe \\
    McI 789 & cut7-24, leu1-32, ura4-d18, h$^{+}$ & I. Hagan \\
  \end{tabular}
  \caption{\footnotesize Fission yeast strains used in this study.}
  \label{table:strain_table}
\end{table*}

\subsection{Cell preparation for imaging}
We grew cells in liquid Edinburgh Minimal Medium (EMM) plus
supplements in an overnight preculture, then diluted them over the
next 24 hours to keep the cells in mid-exponential growth.  Cells were
centrifuged at 2,000 rpm for 5 min on a Beckman CS-6 centrifuge
(Beckman Coulter, Brea, California).  The supernatant was removed and
the cells were suspended in 100 $\mu$L of EMM filtered with a 0.2
$\mu$m filter to reduce background fluorescence from the medium.  The
solution was placed onto 35 mm no.~1.5 glass bottom dishes (MatTek,
Ashland, Massachusetts) coated with 8 $\mu$L of 2 mg/mL lectins from
\textit{Bandeiraea simplicifolia} (Sigma-Aldrich, St. Louis,
Missouri).  The cells were allowed to settle and adhere to the lectins
for 10 min, then the remaining loose cells were washed away twice with
fresh, filtered EMM.  3 mL of fresh, filtered EMM was added to each
dish, and the dishes were placed at 36\degree.  




\section{Results}

\subsection{Sensitivity analysis}

We performed sensitivity analysis to check how the
  capture time and MT mean length vary with model parameters. As
  discussed in the main text and above, dynamic instability with no
  boundary effects gives a mean length of $\meanl = v_g/f_{+0}$ in the
  slow model and
  $\meanl = v_g v_s/(v_s f_c - v_g f_r) \approx v_g/f_c$ in the fast
  model.  To test these relationships, we performed a global
  sensitivity analysis of the mean capture time and MT length to the
  dynamic instability parameters. To quantify this sensitivity, we
  relied on the analysis of the variance of $\ttc$ and $\meanl$ based
  on the so-called Sobol' decomposition \cite{sobol90}, which we
  computed directly using the PC expansion \cite{sudret08}.  The
  dynamic instability parameters were treated as random variables
  uniformly distributed over the ranges in table
  \ref{table:kc_cap_params}. The errors from the PC expansion were
  typically a few percent (ranging from 1-5\% depending on which
  variable and model were fit, table \ref{table:pc_error}). This
  indicates that the PC expansion is accurately capturing the full
  simulation results.
  
\begin{table*}[t]
  \centering
  \begin{tabular}{ p{6 cm}p{1 cm} }
    \textbf{Slow model} &  \textbf{Error} \\
    $\ttc$ for search and capture & 4.9\%\\ 
    $\ttc$ with rotational diffusion & 3.5\% \\
    $\meanl$ for search and capture & 4.4\%\\
    $\meanl$ with rotational diffusion & 0.7\% \\
    \hline
    \textbf{Fast model} &   \\
    $\ttc$ for search and capture &  4.6\%\\ 
    $\ttc$ with rotational diffusion & 3.5\%\\
    $\meanl$ for search and capture &  4.4\%\\
    $\meanl$ with rotational diffusion & 1.1\%\\
  \end{tabular}
  \caption{\footnotesize Error in PC expansion. }
  \label{table:pc_error}
\end{table*}

The overall contribution of each parameter to the
  solution variance was quantified using its total Sobol' index
  \cite{sudret08}; a larger value indicates a higher degree of
  sensitivity of the model to that parameter (table
  \ref{table:sobol_indices}).  As expected, $\ttc$ and $\meanl$ are
  most sensitive to the growth speed and effective catastrophe
  frequency, and this dependence is not altered significantly by the
  addition of rotational diffusion to the model.
  
  \begin{table*}[b]
  \centering
  \begin{tabular}{ p{8 cm}p{1 cm}p{1cm}p{2cm}p{2cm}p{2cm} }
    \textbf{Total Sobol' indices for slow model} &  $v_{g}$ & $f_{+0}$ & $v_{s}$ &
                                                                                   $f_{0-}$
    & $f_{-0}$ \\
    $\ttc$ for search and capture & \textbf{0.74} & \textbf{0.32} &
                                                                    $1.9 \times 10^{-2}$                                                                           &
                                                                                                                                                                     $1.8
                                                                                                                                                                     \times  
                                                                                                                                                                     10^{-2}$&
                                                                                                                                                                               $8.9 \times 10^{-3}$\\ 
    $\ttc$ with rotational diffusion & \textbf{0.70} & \textbf{0.39} &  $9.2 \times 10^{-3}$ &
                                                                                               $1.3
                                                                                               \times
                                                                                               10^{-2}$
    & $8.0
      \times
      10^{-3}$ \\
    $\meanl$ for search and capture & \textbf{0.60} & \textbf{0.42} &  $4.3 \times
                                                                      10^{-3}$ &  $6.5
                                                                                 \times
                                                                                 10^{-3}$
    &  $3.7     \times
      10^{-3}$\\
    $\meanl$ with rotational diffusion & \textbf{0.52} & \textbf{0.50} & $4.0 \times 10^{-4}$ & $2.0
                                                                                                \times
                                                                                                10^{-4}$&$2.0
                                                                                                          \times
                                                                                                          10^{-4}$
    \\
    \hline
    \textbf{Total Sobol' indices for fast model} &  $v_{g}$ & $f_c$ & $v_{s}$ &
                                                                                $f_r$ & \\
    $\ttc$ for search and capture & \textbf{0.85} & \textbf{0.28} & $4.0 \times 10^{-3}$
                                                                                 &
                                                                                   $7.5
                                                                                   \times  
                                                                                   10^{-3}$& \\ 
    $\ttc$ with rotational diffusion & \textbf{0.84} & \textbf{0.28} &  $7.3 \times 10^{-3}$ &
                                                                                               $9.9
                                                                                               \times
                                                                                               10^{-3}$ & \\
    $\meanl$ for search and capture & \textbf{0.49} & \textbf{0.39} &  $8.1 \times
                                                                      10^{-2}$ &  $9.5
                                                                                 \times
                                                                                 10^{-2}$
    & \\
    $\meanl$ with rotational diffusion & \textbf{0.50} & \textbf{0.39} & $7.6 \times 10^{-2}$ & $9.7
                                                                                                \times
                                                                                                10^{-2}$& \\
  \end{tabular}
  \caption{\footnotesize Total Sobol' indices. }
  \label{table:sobol_indices}
\end{table*}

\begin{figure*}[t]
  \centering
  \includegraphics[width=1.0 \textwidth]{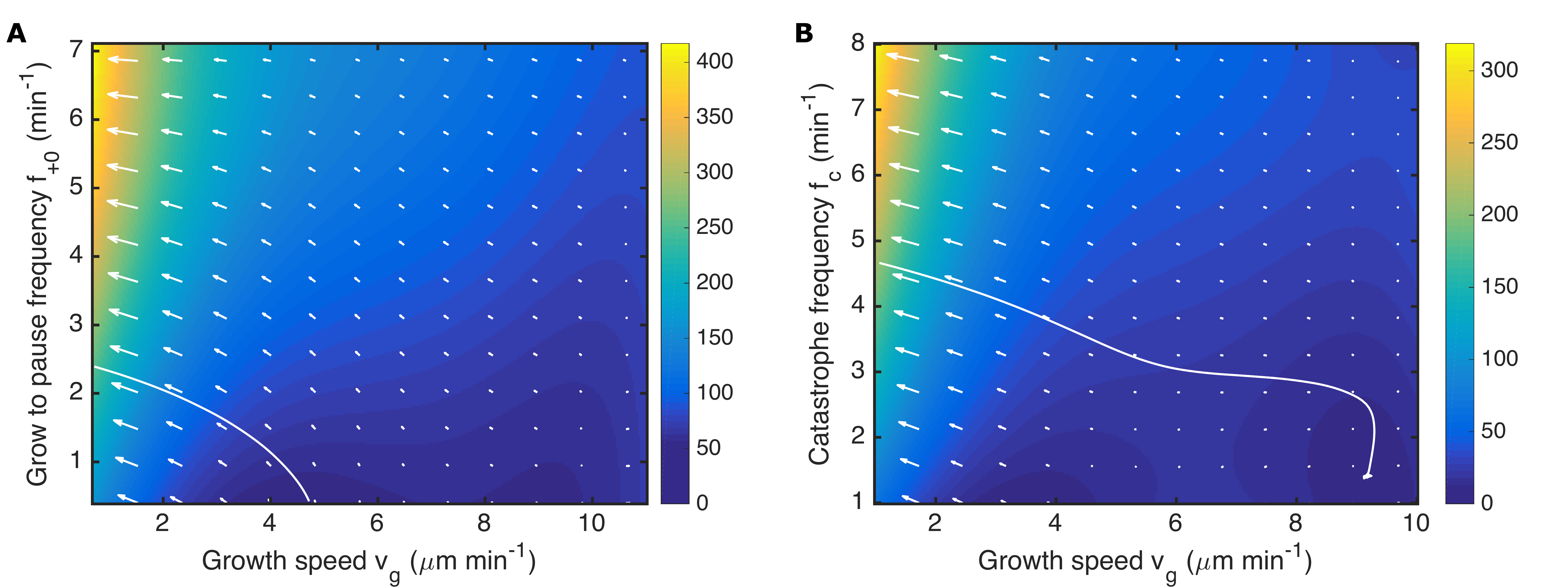}
  \caption {\footnotesize Dependence of capture time
      on growth speed and effective catastrophe frequency for search
      and capture model determined from polynomial chaos
      expansion. (A) Slow model with no MT rotational diffusion. (B)
      Fast model with no MT rotational diffusion.  Color shows capture
      time in minutes.  White arrows show the magnitude and direction
      of most rapid change in $\ttc$ (the gradient). The white line is
      the curve that follows the gradient and passes through the
      reference parameter set.  }
  \label{fig:ttc_vs_vg_fcat_fixed}
\end{figure*}

We also determined the local sensitivity of the
  capture time to these parameters by computing the gradient (the
  local direction of steepest change) of $\ttc$ as a function of the
  growth speed and catastrophe frequency from the PC expansion, with
  other parameters fixed at their reference values
  (fig. \ref{fig:ttc_vs_vg_fcat_fixed} shows results for the models
  without MT rotational diffusion; here the white line follows the
  local gradient and passes through the reference parameters). This
  local sensitivity analysis illustrates the same result found for the
  global analysis: the capture time is primarily controlled by the
  growth speed (primarily horizontal arrows in
  fig. \ref{fig:ttc_vs_vg_fcat_fixed}) with the main secondary effect
  from the catastrophe frequency (vertical component of arrows in
  fig. \ref{fig:ttc_vs_vg_fcat_fixed}). It also illustrates that the
  shortest capture times occur for high growth speed and low
  catastrophe frequency where MT mean lengths are longest.

\begin{figure*}
  \centering

  \includegraphics[width=0.32 \textwidth]{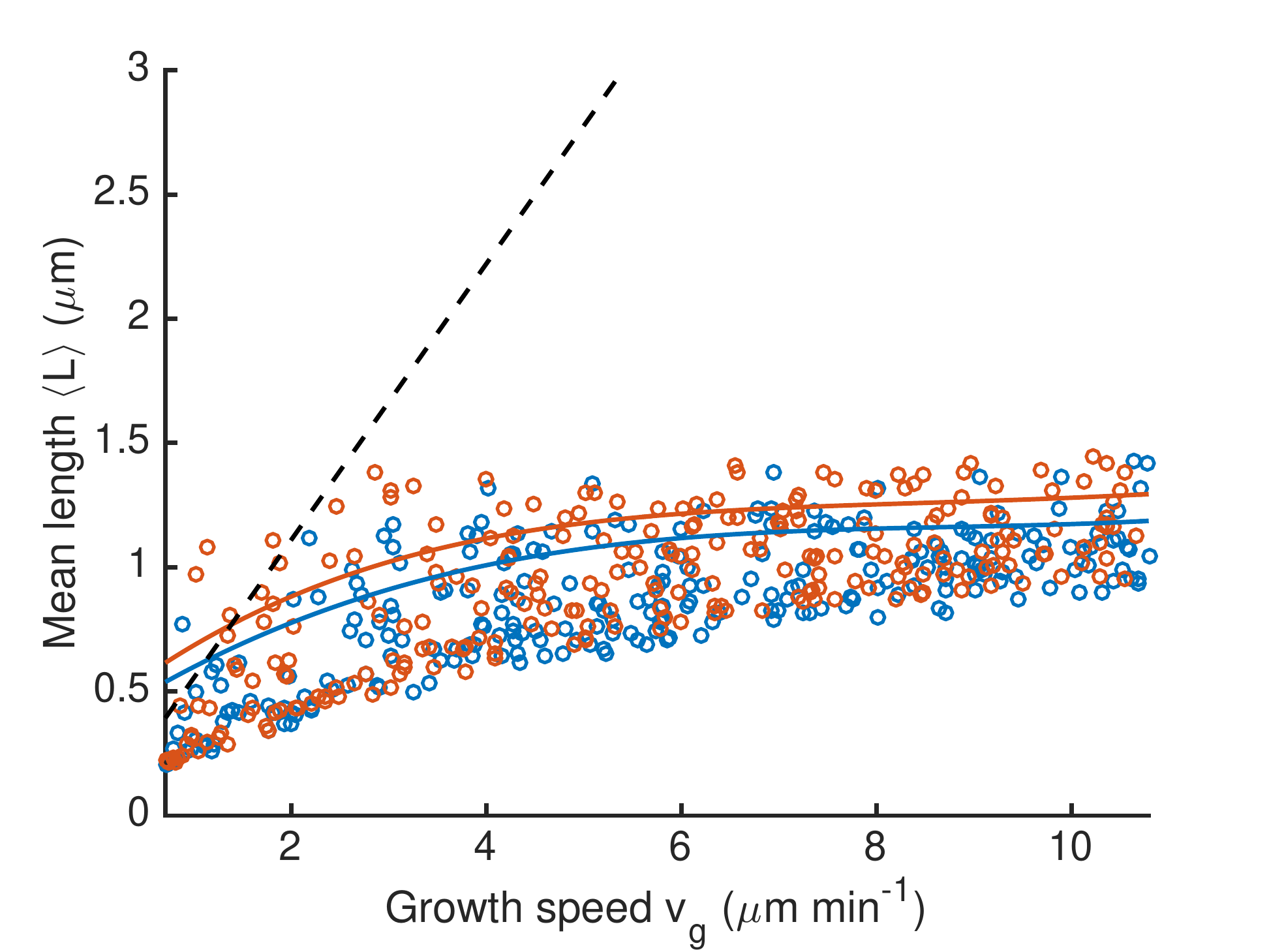}
  \includegraphics[width=0.32 \textwidth]{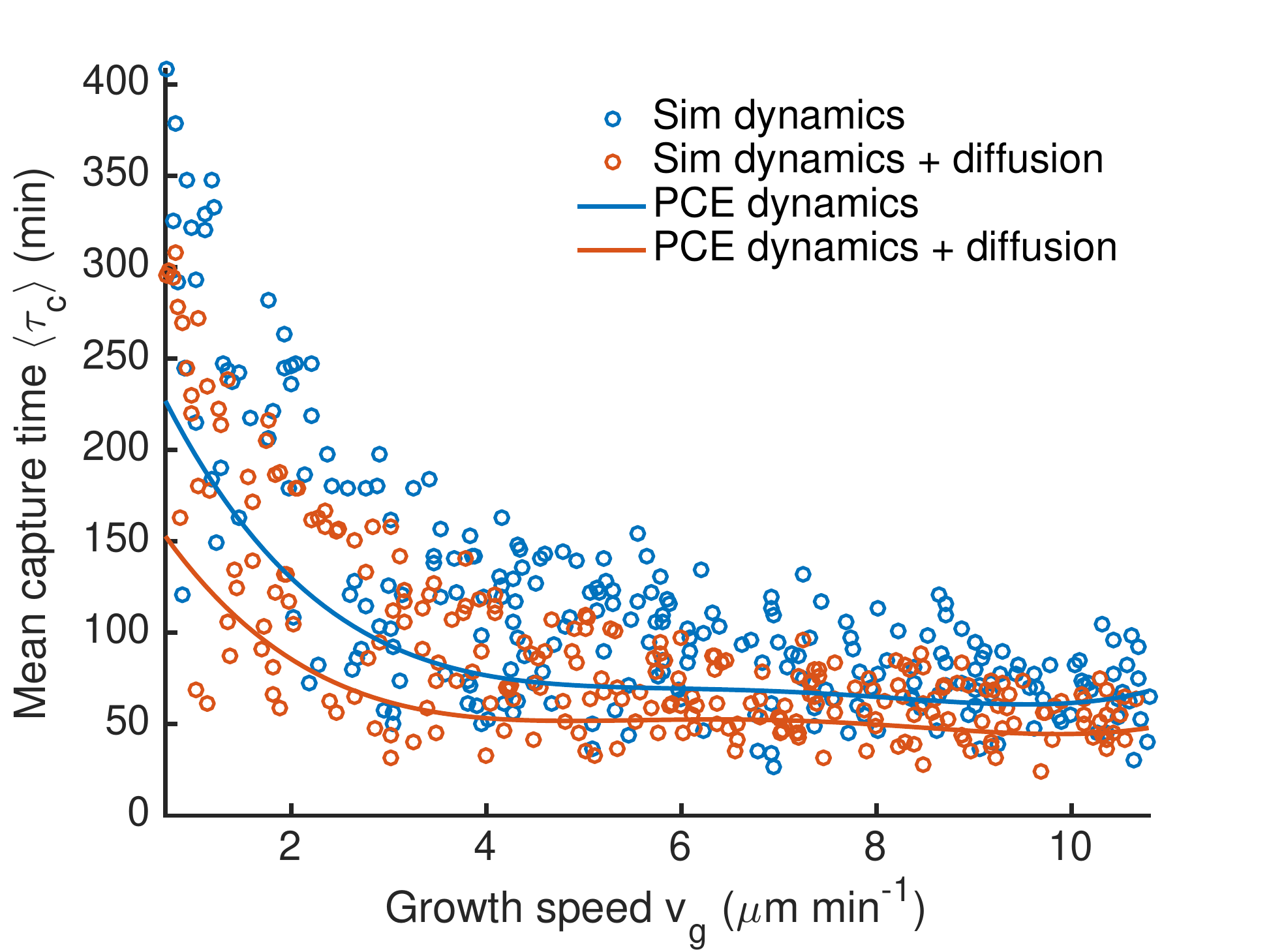}\\
  \includegraphics[width=0.32 \textwidth]{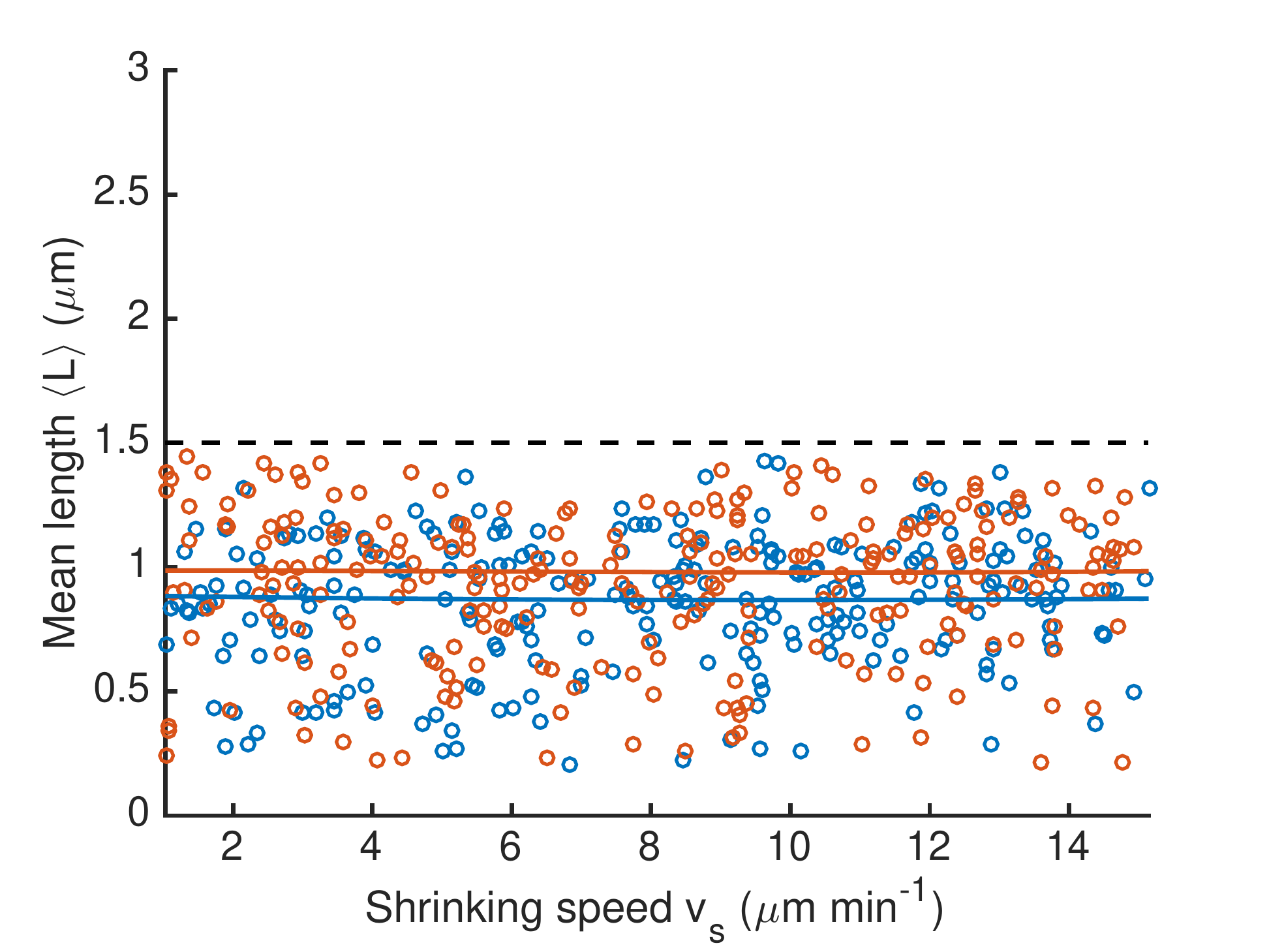}
  \includegraphics[width=0.32 \textwidth]{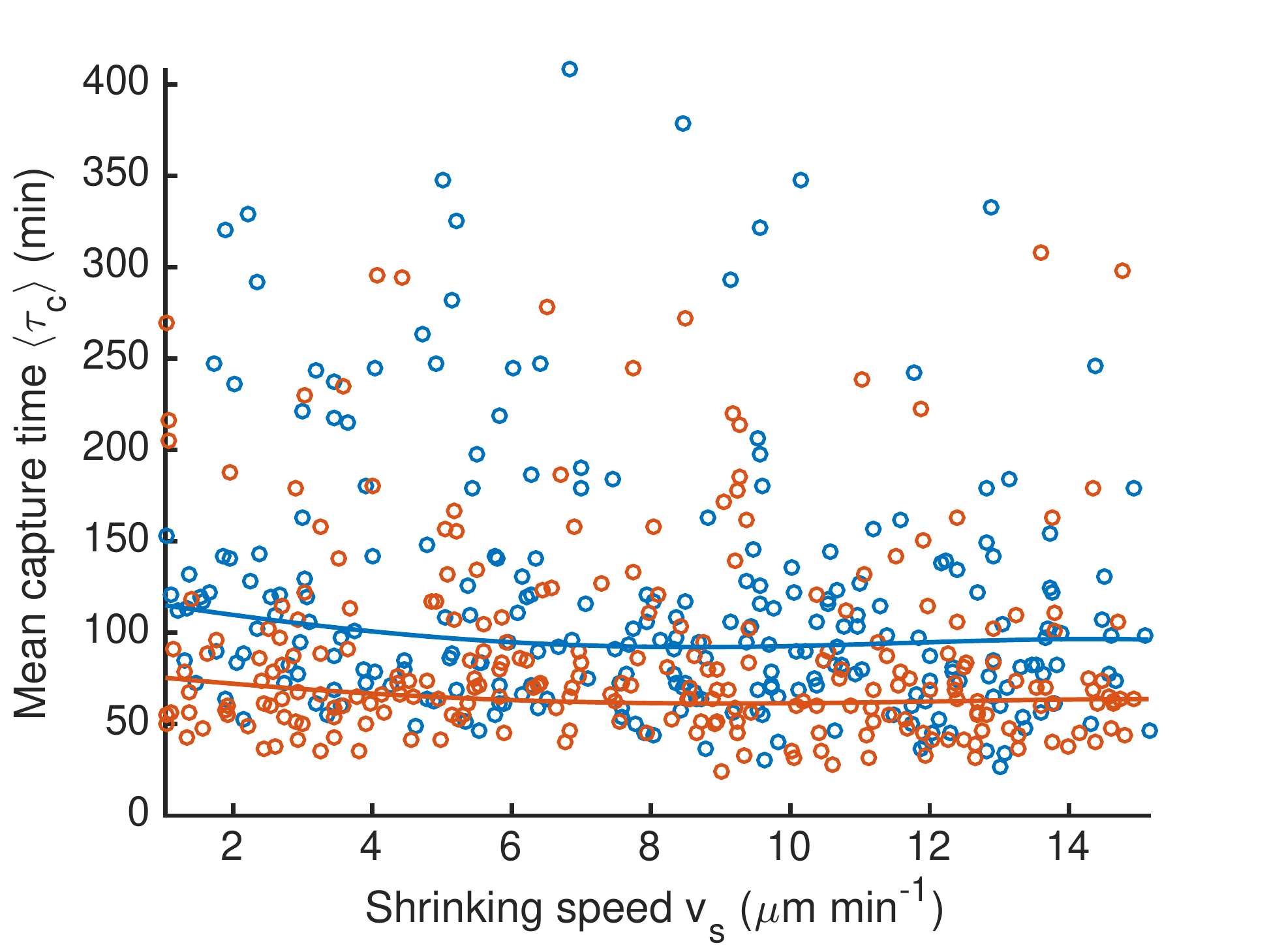}\\
  \includegraphics[width=0.32 \textwidth]{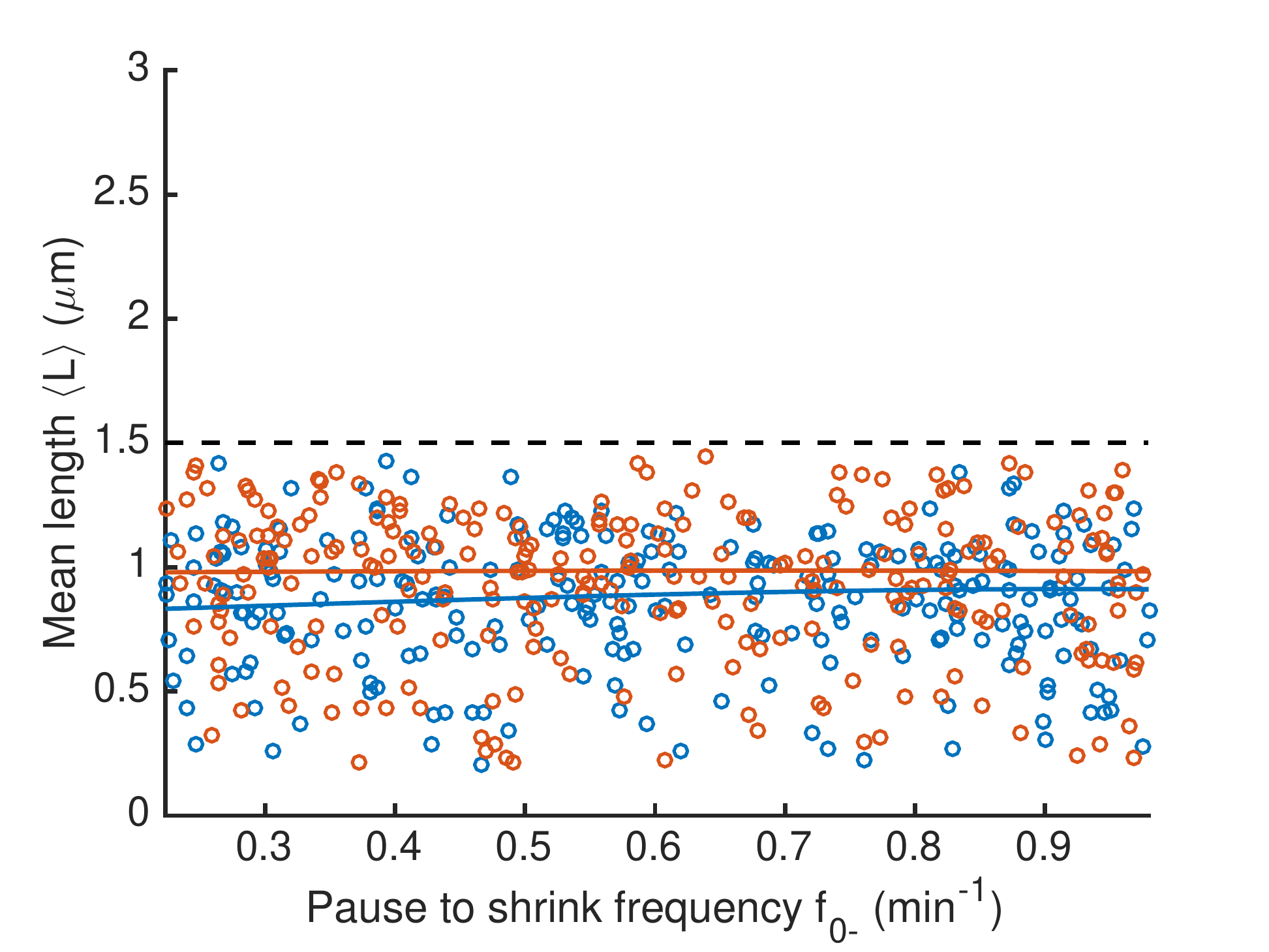}
  \includegraphics[width=0.32 \textwidth]{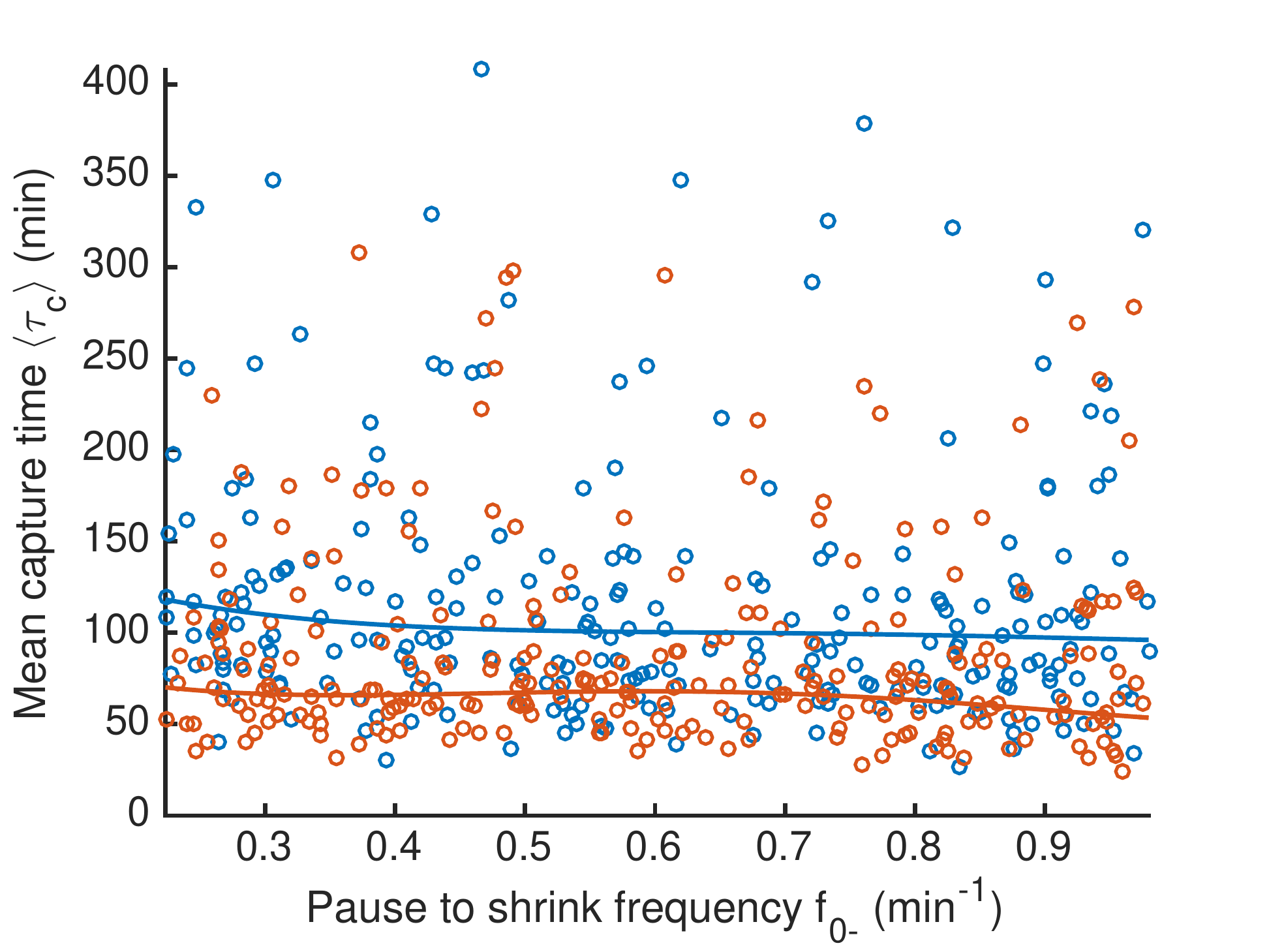}\\
  \includegraphics[width=0.32 \textwidth]{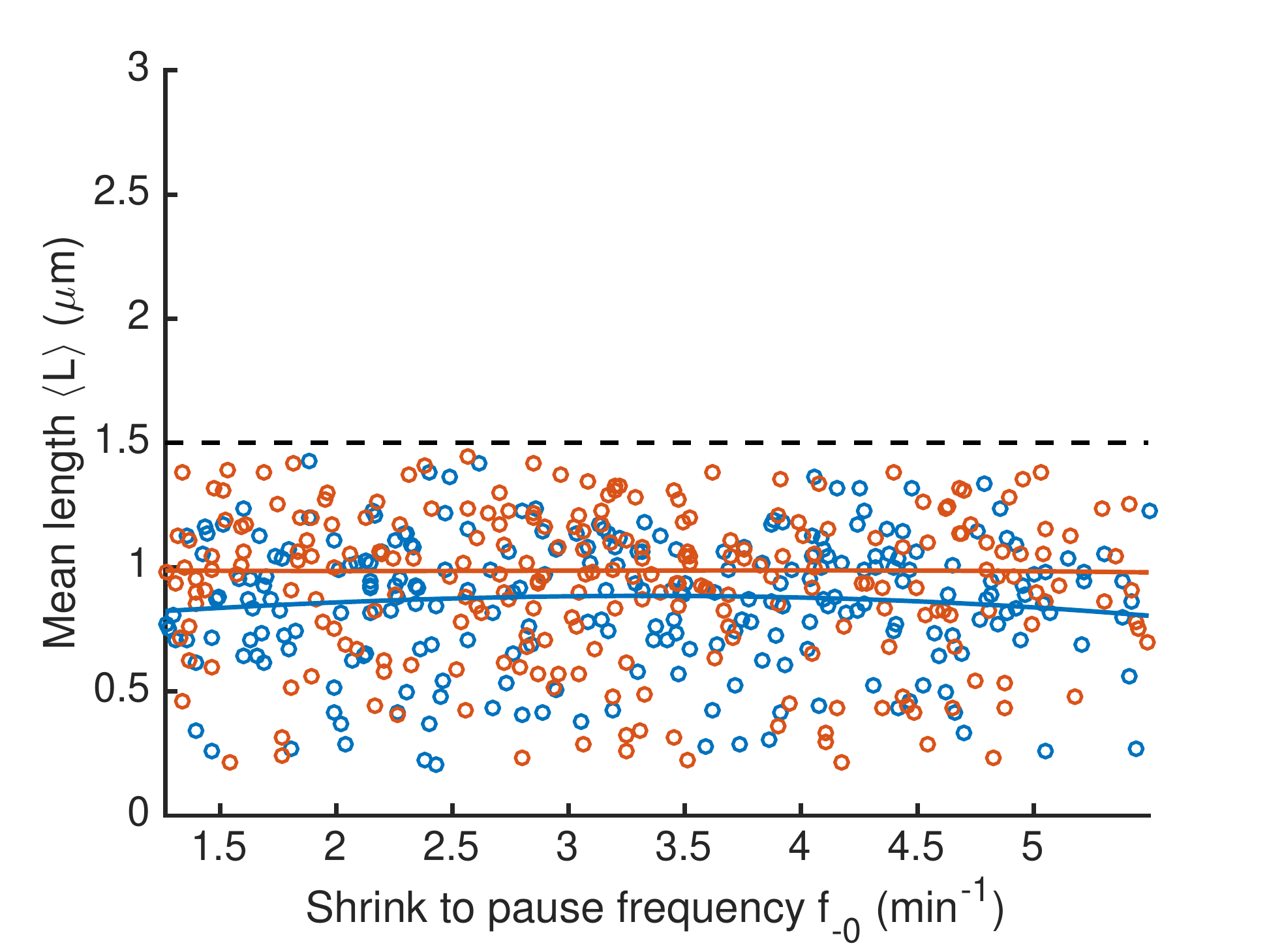}
  \includegraphics[width=0.32 \textwidth]{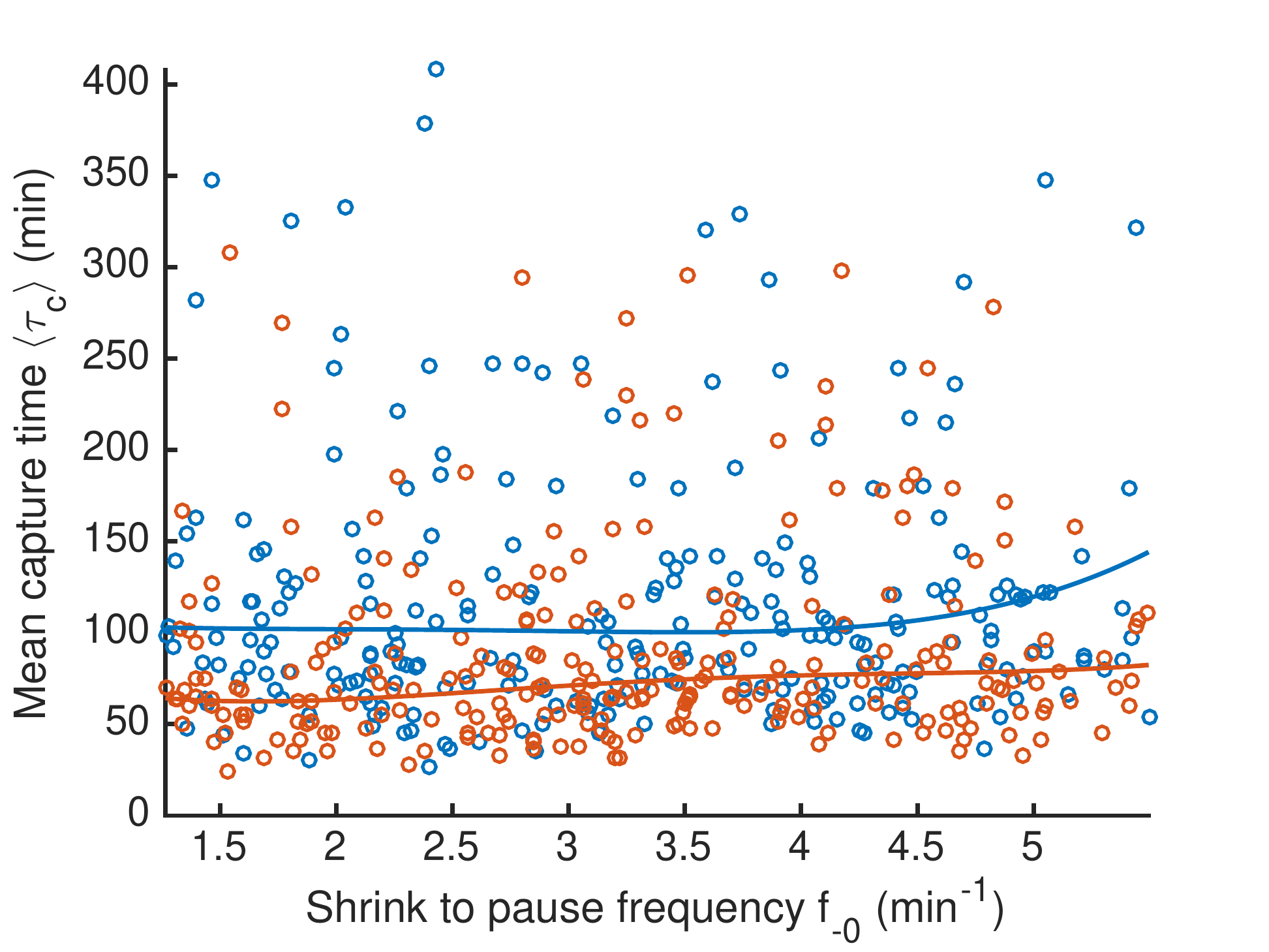}\\
  \includegraphics[width=0.32 \textwidth]{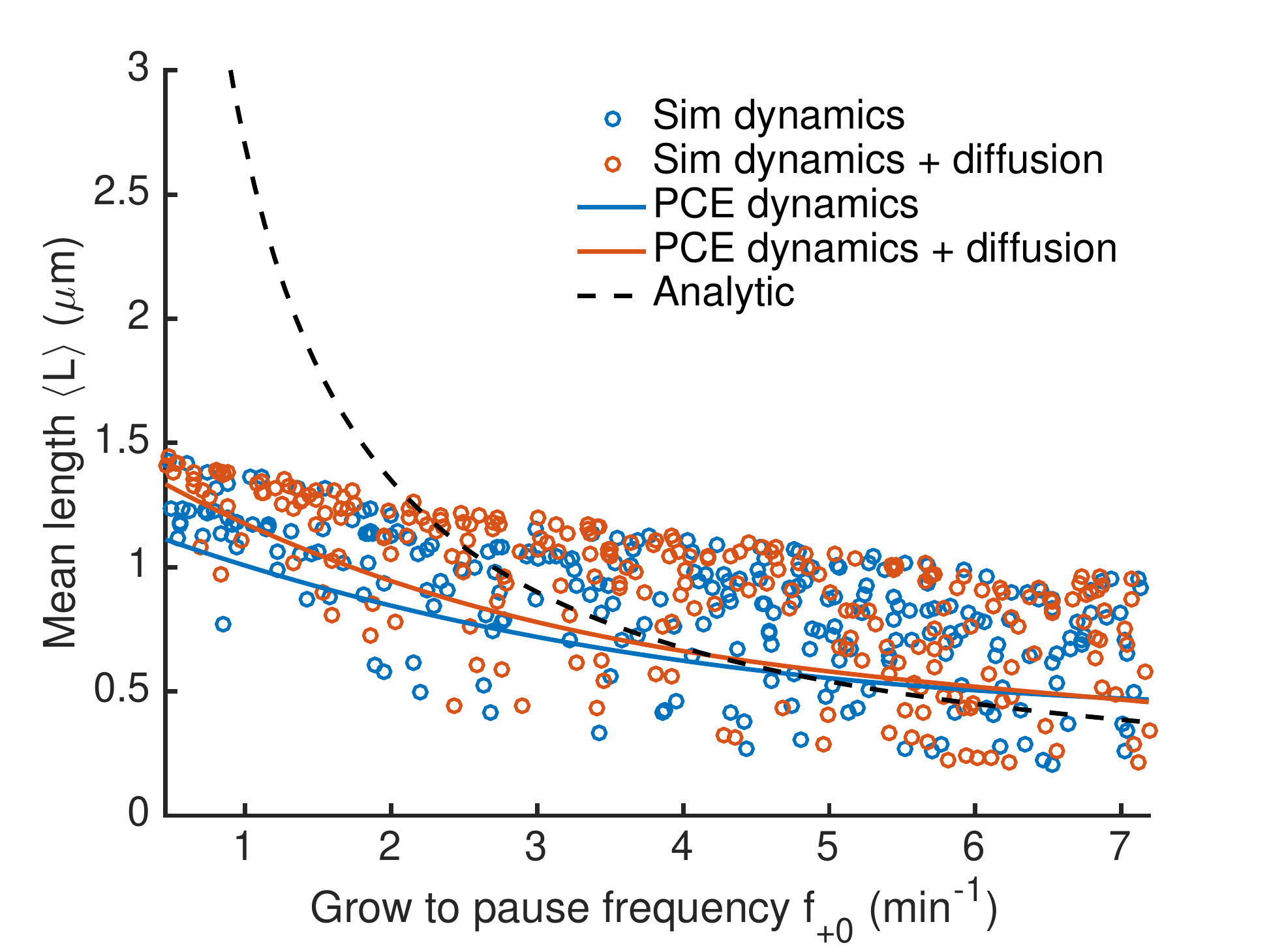}
  \includegraphics[width=0.32 \textwidth]{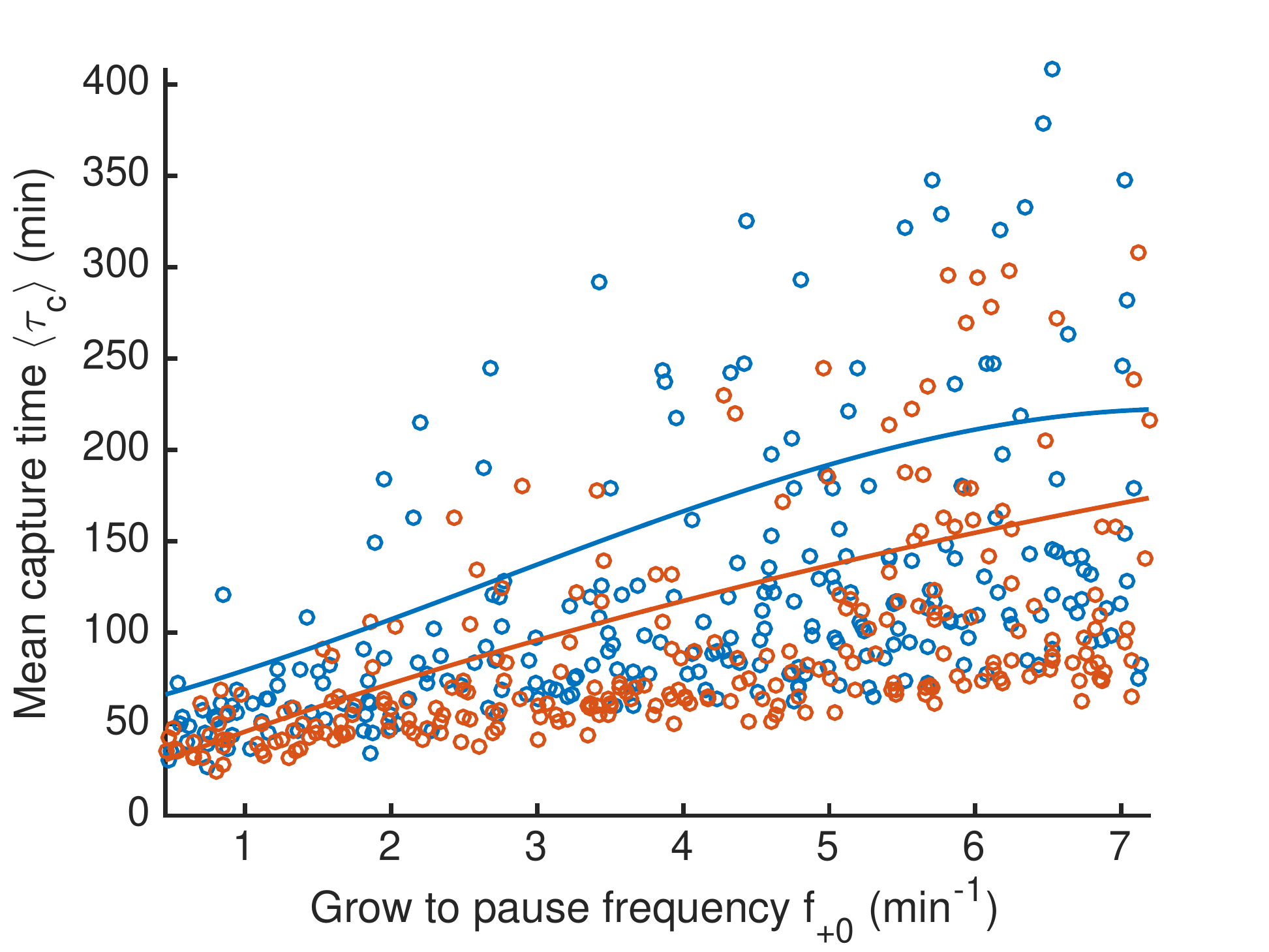}
  \caption {\footnotesize $\meanl$ and $\ttc$ vs. parameters for slow
    model.   \label{suppfig:l_ttc_vs_all_slow}}
\end{figure*}

\begin{figure*}
  \centering
  \includegraphics[width=0.4 \textwidth]{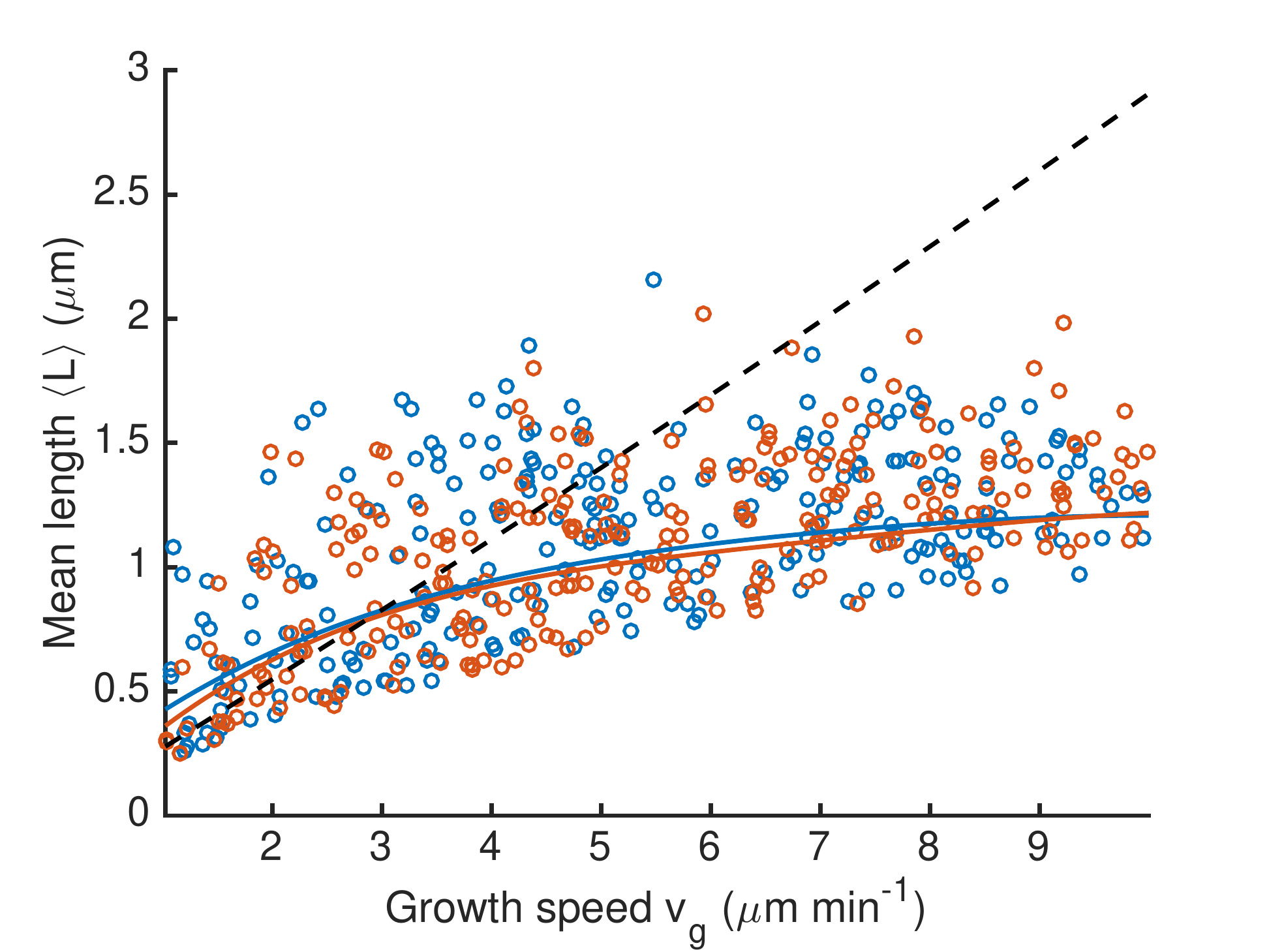}
  \includegraphics[width=0.4 \textwidth]{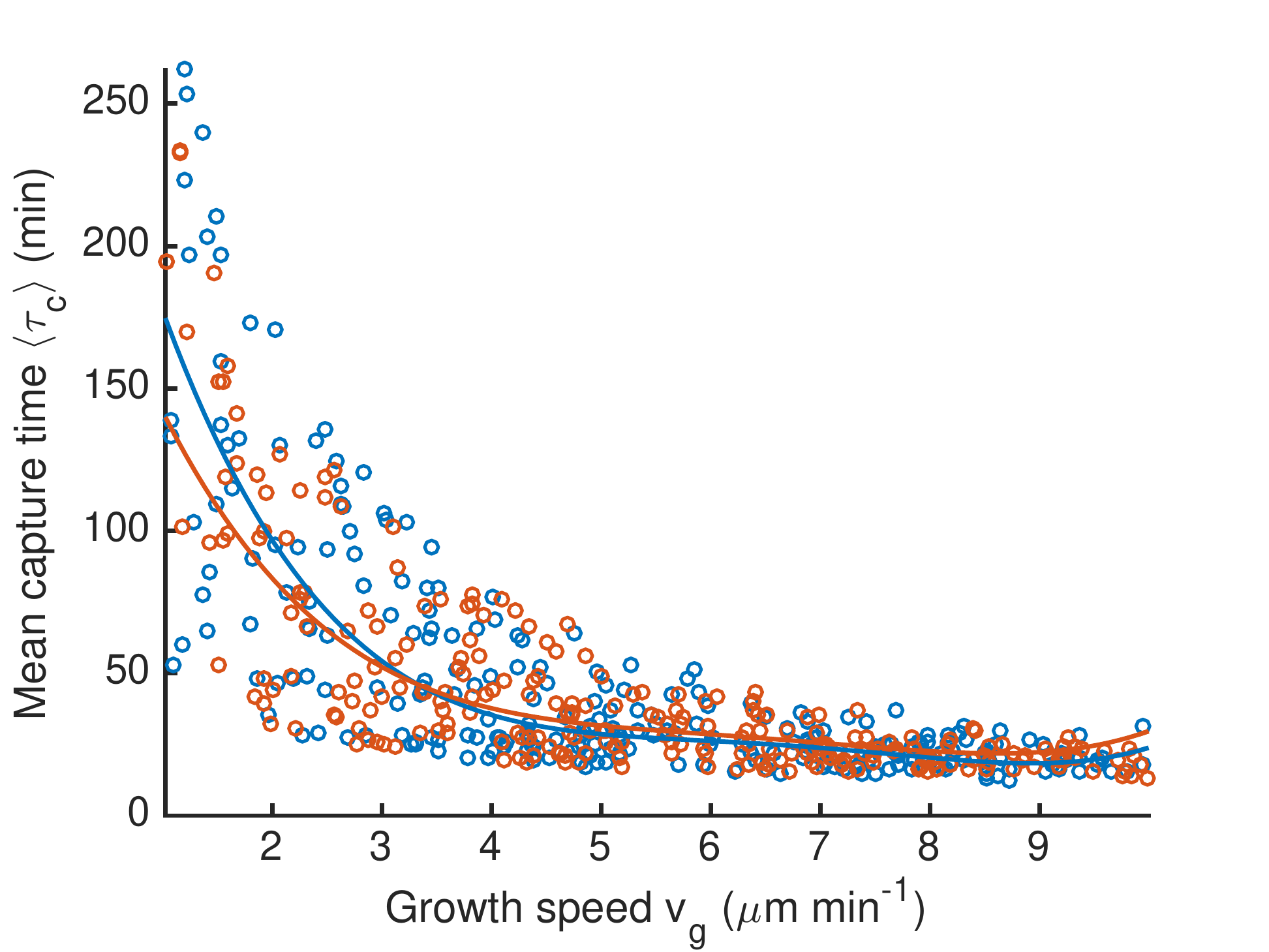}
  \includegraphics[width=0.4 \textwidth]{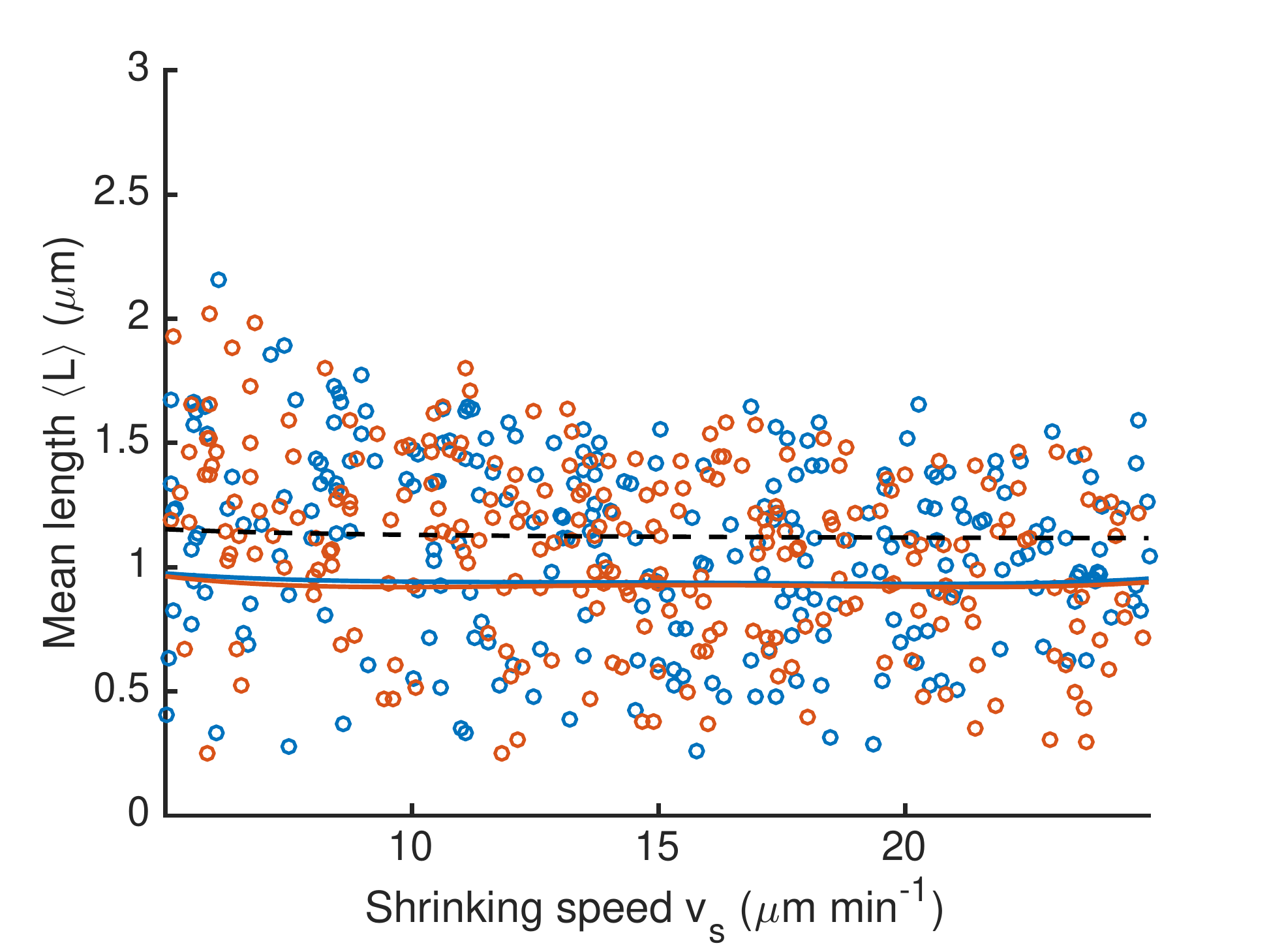}
  \includegraphics[width=0.4 \textwidth]{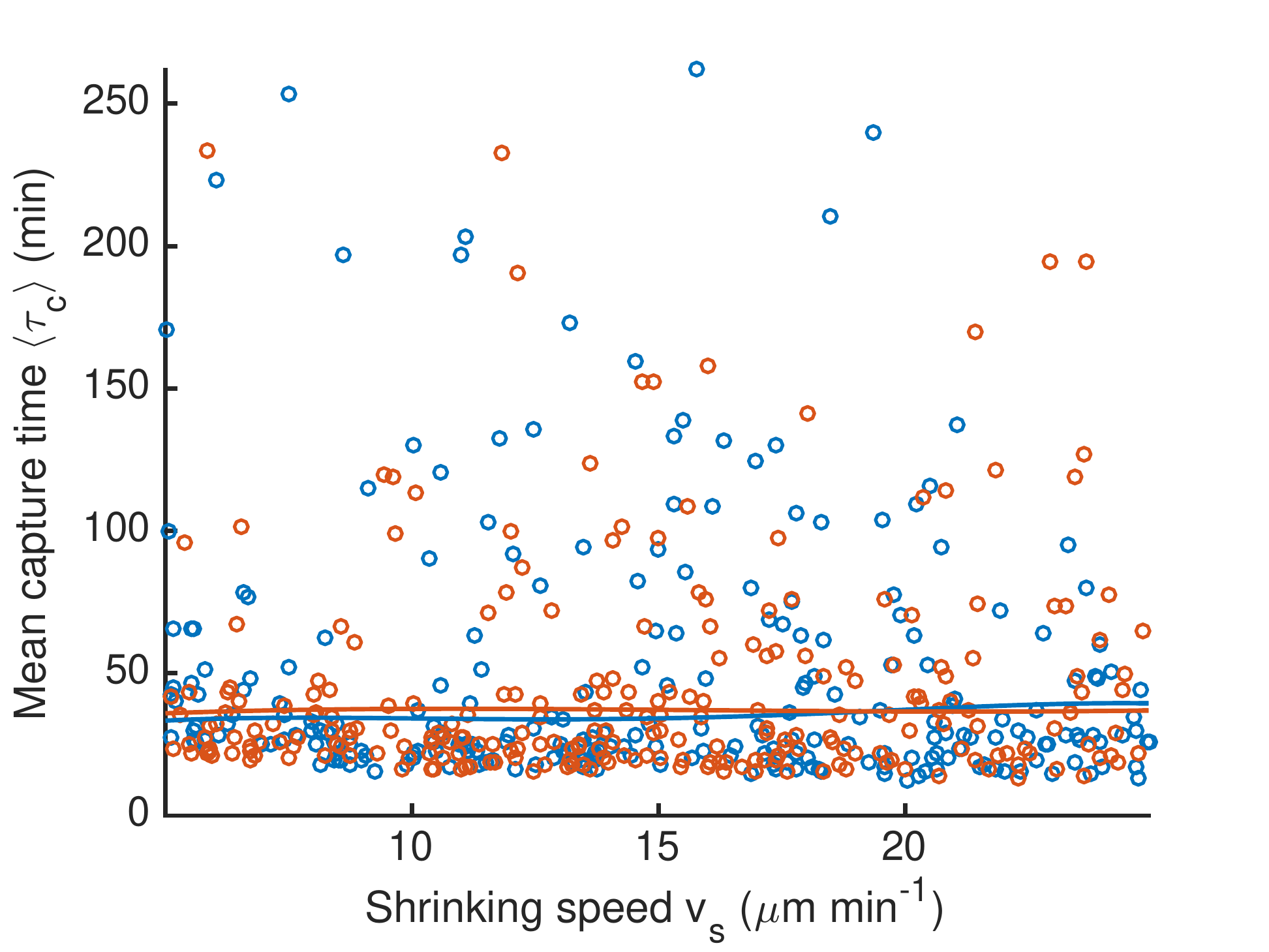}
  \includegraphics[width=0.4 \textwidth]{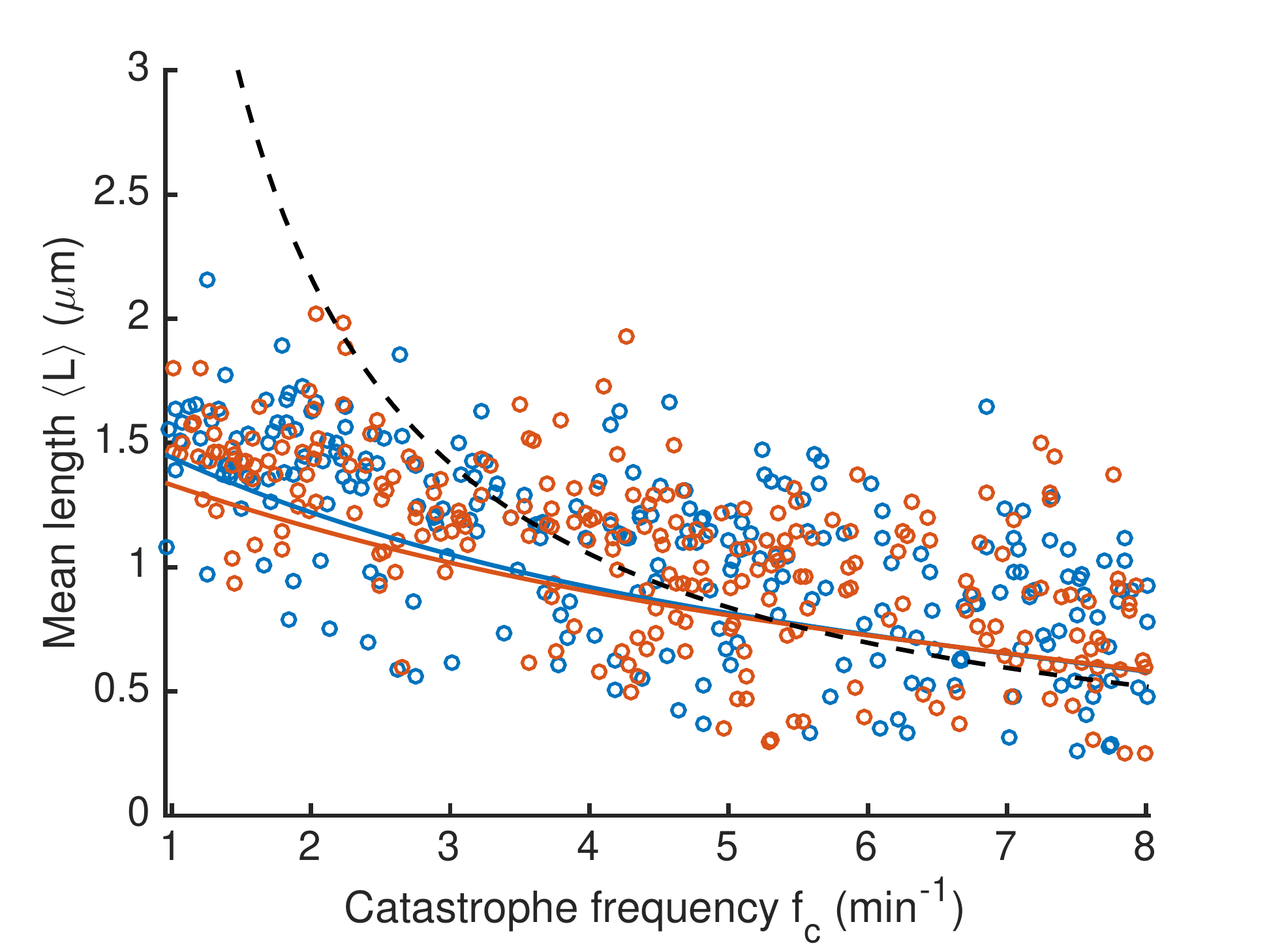}
  \includegraphics[width=0.4 \textwidth]{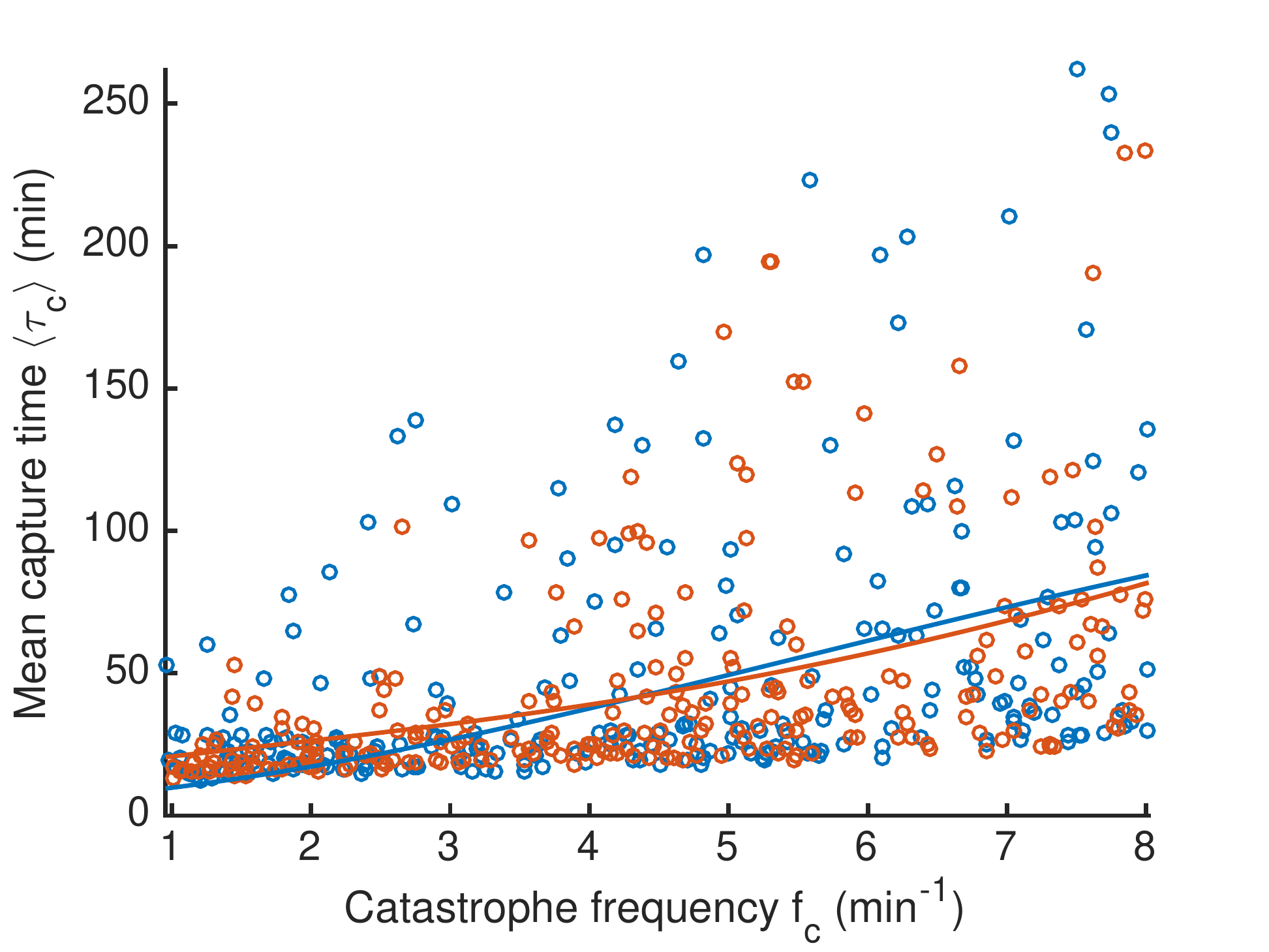}
  \includegraphics[width=0.4 \textwidth]{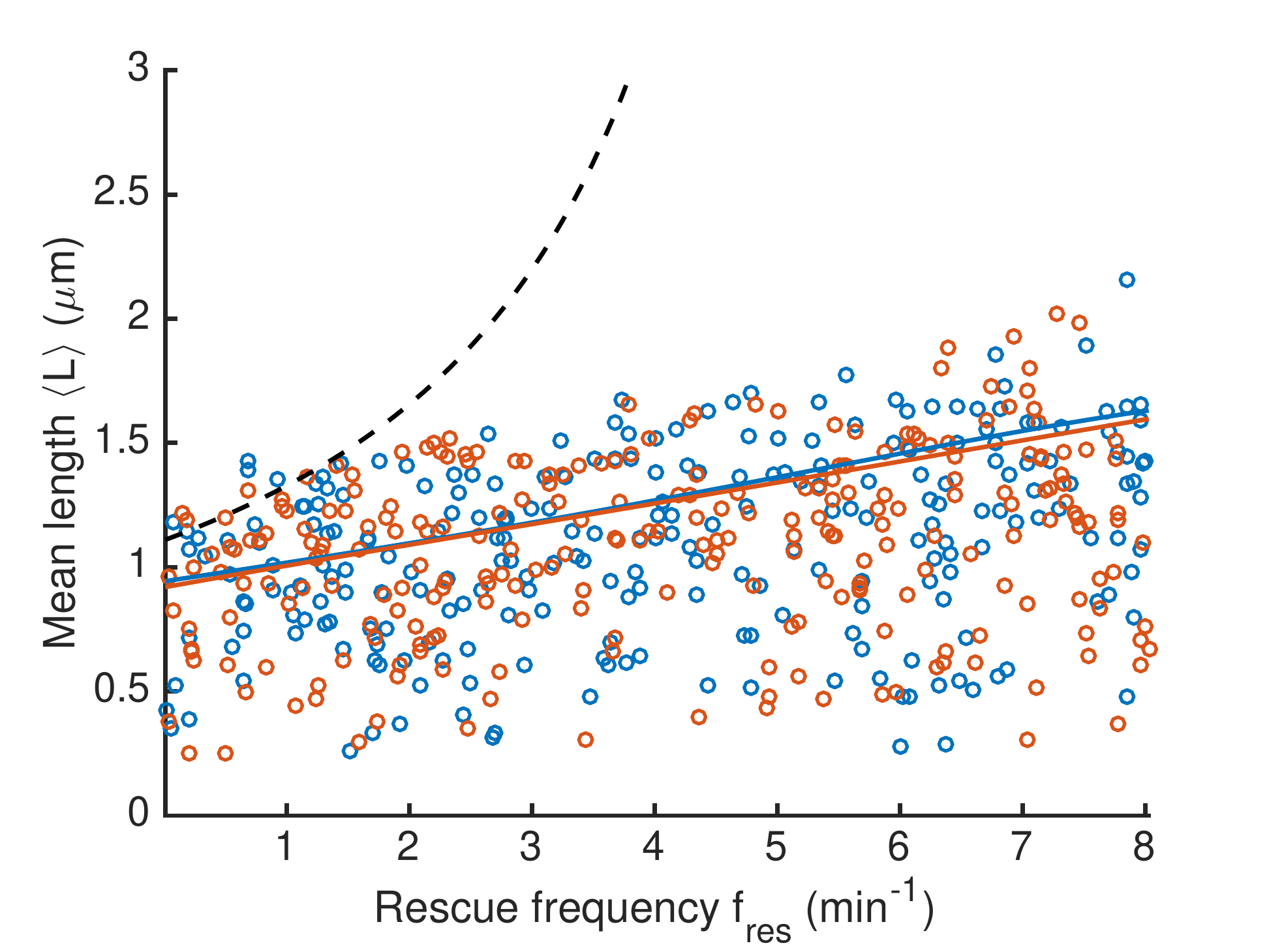}
  \includegraphics[width=0.4 \textwidth]{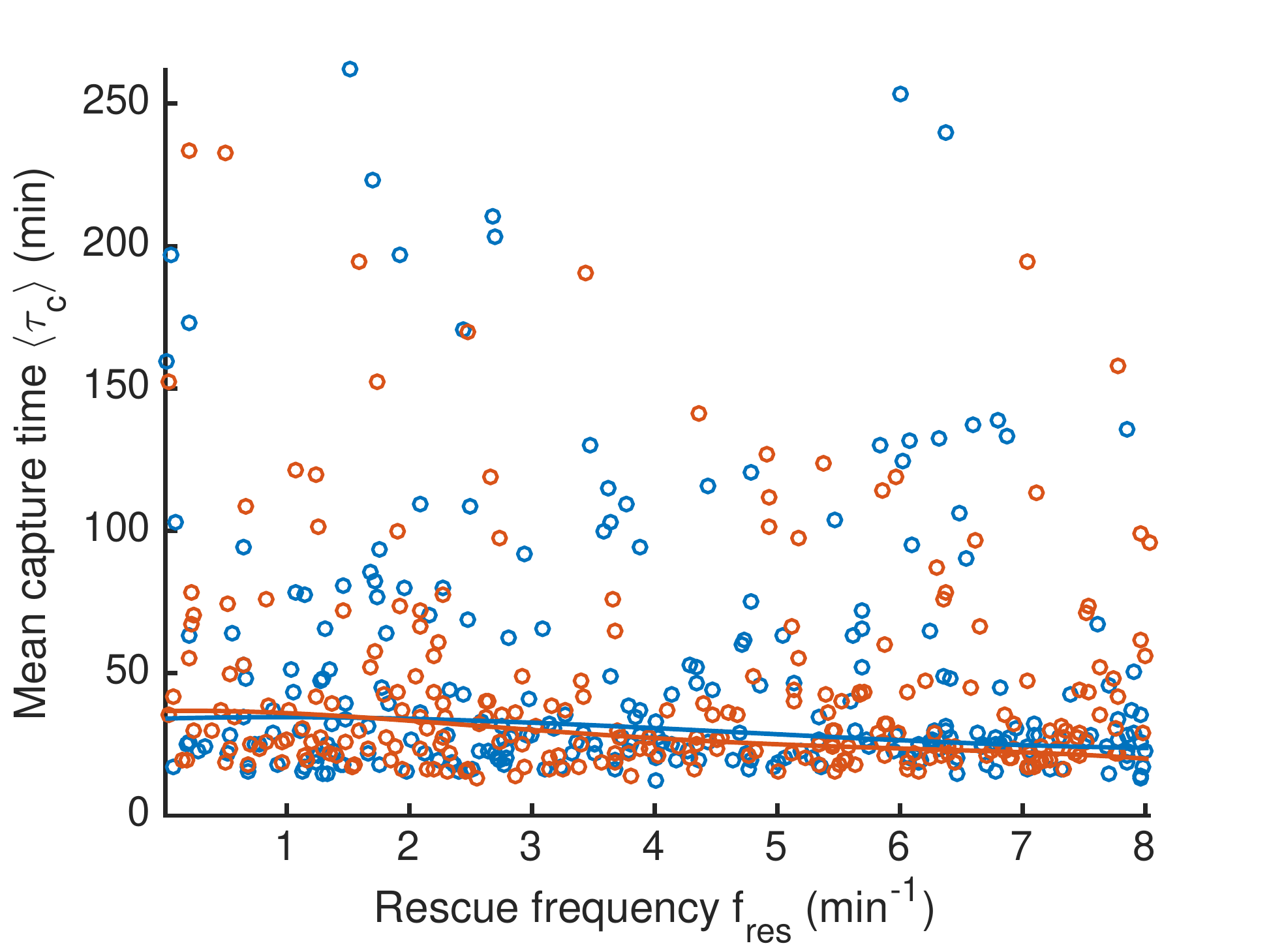}
  \caption {\footnotesize $\meanl$ and $\ttc$ vs. parameters for fast
    model.   \label{suppfig:l_ttc_vs_all_fast}}
\end{figure*}


\end{document}